\documentclass{aa} 
\usepackage{graphicx}
\usepackage{txfonts}
\usepackage{natbib}
\usepackage{amsmath}
\usepackage{color}
\usepackage{array}
\usepackage{multirow}
\usepackage{subfig}
\usepackage{xspace}
\usepackage{xcolor}
\usepackage{url}
\usepackage{rotating}
\usepackage{pdflscape}

\newcommand{\equ}[1]{Eq.~\ref{eq:#1}}

\newcommand{\fig}[1]{Fig.~\ref{fig:#1}}

\newcommand{\sect}[1]{Sect.~\ref{sec:#1}}
\newcommand{\app}[1]{Appendix~\ref{app:#1}}

\newcommand{\atlas}[0]{ATLAS$^\mathrm{3D}$\xspace}
\newcommand{\lcdm}[0]{$\Lambda$CDM\xspace}
\newcommand{\rph}[0]{\ensuremath{\rho_\mathrm{ph}}\xspace}
\newcommand{\rphtilde}[0]{\ensuremath{\widetilde{\rho}_\mathrm{ph}}\xspace}
\newcommand{\sigph}[0]{\ensuremath{\Sigma_\mathrm{ph}}\xspace}
\newcommand{\sigphtilde}[0]{\ensuremath{\widetilde{\Sigma}_\mathrm{ph}}\xspace}
\newcommand{\sigcr}[0]{\ensuremath{\Sigma_\mathrm{cr}}\xspace}
\newcommand{\rM}[0]{\ensuremath{r_\mathrm{M}}\xspace}
\newcommand{\req}[0]{\ensuremath{r_\mathrm{ef}}\xspace}
\newcommand{\reff}[0]{\ensuremath{r_\mathrm{e}}\xspace}
\newcommand{\mtot}[0]{\ensuremath{M_\mathrm{tot}}\xspace}
\newcommand{\rmax}[0]{\ensuremath{r_\mathrm{max}}\xspace}
\newcommand{\zmax}[0]{\ensuremath{z_\mathrm{max}}\xspace}
\newcommand{\zmin}[0]{\ensuremath{z_\mathrm{min}}\xspace}
\newcommand{\mneg}[0]{\ensuremath{M_\mathrm{neg}}\xspace}

\begin{document}

\title{Peculiar dark matter halos inferred from gravitational lensing as a manifestation of modified gravity}

\author{Michal B\'{i}lek\inst{1,2,3}
}
\institute{Observatoire de Paris, LERMA, Coll\`ege de France, CNRS, PSL University, Sorbonne University, F-75014, Paris\\
    \email{michal.bilek@obspm.fr}
    \and
    Strasbourg University, CNRS, Observatoire astronomique de Strasbourg, F-67000 Strasbourg, France
    \and FZU – Institute of Physics of the Czech Academy of Sciences, Na Slovance
1999/2, Prague 182 21, Czech Republic
}

\date{Received ...; accepted ...}

\abstract{If modified gravity holds, but the weak lensing analysis is done in the standard way, one finds that dark matter halos have peculiar shapes, not following the standard Navarro-Frenk-White profiles, and are fully predictable from the distribution of baryons. Here we study in detail the distribution of the apparent dark matter around point masses, which approximate galaxies and galaxy clusters, and their pairs for the QUMOND MOND gravity, taking an external gravitational acceleration $g_e$ into account. At large radii, the apparent halo of a point mass $M$ is shifted against the direction of the external field. When averaged over all lines-of-sight, the halo has a hollow center, and denoting the by $a_0$ the MOND acceleration constant, its density behaves like $\rho(r) = \sqrt{Ma_0/G}/(4\pi r^2)$ between the galacticentric radii $\sqrt{GM/a_0}$ and $\sqrt{GMa_0}/g_e$, and like $\rho\propto r^{-7}G^2M^3a_0^3/g_e^5$ further away. Between a pair of point masses, there is a region of a negative apparent dark matter density, whose mass can exceed the baryonic mass of the system. The density of the combined dark matter halo is not a sum of the densities of the halos of the individual points. The density has a singularity near the zero-acceleration point, but remains finite in projection. We compute maps of the surface density and the lensing shear for several configurations of the problem, and derive formulas to scale them to further configurations. In general, for a large subset of MOND theories in their weak field regime, for any configuration of the baryonic mass $M$ with the characteristic size of $d$, the total lensing density scales as $\rho({\vec{x}}) = \sqrt{Ma_0/G}d^{-2}f\left(\vec{\alpha}, \vec{x}/d, g_ed/\sqrt{GMa_0}\right)$, where the vector $\vec{\alpha}$ describes the geometry of the system. Detecting the difference between QUMOND and cold dark matter halos appears to be possible with the existing instruments.}


\maketitle

\section{Introduction} \label{sec:intro}
The missing mass problem has not been solved decisively yet.  The usual solution is to assume the validity of Newtonian dynamics and general relativity and to postulate the existence of dark matter (see, e.g., \citealp{salucci19} for a review). While many types of dark matter have been proposed, here we for concreteness focus on the cold dark matter and the \lcdm cosmological model. In the \lcdm model, many well-known observations, particularly on the cosmological scales, can be explained well (cosmic microwave background, baryon acoustic oscillations, distance-luminosity relation for type II supernovae, etc.). However, many questions remain unexplained, such as the mismatch between the Hubble constant in the early and late Universe \citep{verde19},  large-scale cosmic flows \citep{peery18},  emptiness of the Local Void \citep{peebles10,haslbauer20}, existence of the planes of satellites \citep{pawlowski19}, among others. The internal properties of galaxies cannot be predicted well from the first principles and are even hard to reproduce in simulations \citep{brooks16,oman15,bullock17}, while simple, tight scaling relations between the baryonic and dark matter content are observed  \citep{famaey12,mcgaugh16}.

Another option is to explain the missing mass problem without dark matter particles outside of the Standard model of particle physics. This is the case of the MOND \citep{milg83a} paradigm, according to which our knowledge of  the law of gravity and/or inertia should be corrected in the limit of small accelerations (see, e.g., the reviews \citealp{milgcjp, milgscholar,famaey12}).
Hereafter in this introduction, we will consider only the MOND modified gravity theories. Although MOND was initially devised on the basis of the first rough observations of rotation curves, it provides formulas that allow predicting dynamics of a wide range of galaxies, while the most reliable measurements show the best agreement \citep{lelli17,li18}.  Importantly for the present paper, MOND also has been tested successfully on the scale of hundreds and thousands of kpc via weak gravitational lensing \citep{milg13,mistele23,mistele24} and the dynamics of galaxy groups \citep{milg18,milg19}. The open questions for MOND include the nature of the remaining missing mass problem of galaxy clusters \citep{milg08}, the survivability and dynamics of ultra-diffuse galaxies in these environments \citep{milgudgs,freundlich22}, the internal dynamics of globular clusters \citep{ibata13b, hernandez20}, and finally finding a covariant version of MOND which would simultaneously explain all the available data  \citep{famaey12,mistele23b}. While MOND might not be the definitive solution to the missing mass problem, the above results show that if a modified gravity theory is governing our Universe, then MOND must be its excellent approximation on the scale of galaxies and galaxy groups.

A key aspect of MOND, absent in the Newtonian gravity, is external field effect (EFE). It is a manifestation of the non-linearity of the MOND theories -- in MOND, the superposition principle for gravitational fields does not hold true \citep{milgmondlaws}.  In reality, galaxies are always embedded in some external field that is generated by nearby galaxies, galaxy clusters and the large-scale cosmological structures. The external field is, of course, different for each object, but has been estimated to be typically on the order of a few percent of $a_0$ \citep{famaey07,wu08,hees11,milg18,oria21} but it can be much higher, for example, for satellites of galaxies or for galaxies in clusters \citep{muller19,freundlich22}.

In modified gravity theories which do not involve dark matter, we introduce the concept of ``phantom dark matter'' (PDM,  \citealp{milg86}). It is a mathematical construct that denotes the dark matter that would lead in the Newtonian gravity to the observed gravitational field while actually a modified gravity theory holds true. The distribution of PDM can be very unusual or even impossible from the point of view of the Newtonian gravity. For example, in MOND the PDM around a point mass has a zero density in the vicinity of the point, but at larger radii it mimics an isothermal halo. This is in stark contrast to the Navarro-Frenk-White dark matter halo expected by the theory of \lcdm, which has an infinite density in its center \citep{nfw}. For some configurations of baryonic matter, the density of PDM can even be negative in some regions of space. \citet{milg86} identified a few specific mass configurations that give rise to negative density of PDM: a pair of point masses, a disk with a hole in its center, an inhomogeneous thin galactic disk, and an infinitesimal mass embedded in a homogeneous external gravitational field.   \citet{oria21} made maps of the regions of the negative PDM for the Local Volume. The distribution of PDM around an object depends, in general, on the  external field imposed on it by other objects. Also, it can strongly depend on the particular MOND theory \citep{gqumond}. 

It was proposed to distinguish modified gravity from dark matter by the EFE \citep{milg83a}, astrophysical tests based on dynamical friction  \citep{kroupacjp}, the dynamics of wide binary stars \citep{hernandez12}, and others. Nevertheless, all these tests have their disadvantages. Another possibility to distinguish between various solutions to the missing mass problem would be to inspect the halos, real or phantom, of galaxies, their groups, and clusters by weak gravitational lensing. All these objects are expected to have virialized dark matter halos in the theories involving particle dark matter. Their detailed profiles depend on the type of the dark matter particle. They can however never take the shapes of some of the PDM halos, such as having negative densities. 

In all relativistic MOND theories known so far, calculating gravitational lensing is not particularly difficult: we can apply the procedures known from general relativity on the sum of the densities of baryonic matter and of the PDM \citep{milg13}.  This has been utilized for explaining some observations. For example, \citet{jee07} reported the discovery of a ring of dark matter around the galaxy cluster  Cl~0024+17. \citet{milg08} showed that such rings are expected to be detected if MOND works but Newtonian gravity is assumed for the reconstruction of the surface density map, since sufficiently compact objects generate PDM halos with central cavities. Establishing if such rings of dark matter are common will be a strong test of MOND. \citet{milg13} showed that the observed halos of galaxies follow MOND predictions. There were even hints that the measured dark halos are cut off at about 200\,kpc. In MOND this is explained, as we detail in \sect{pm}, by the EFE. \citet{mistele23} demonstrated that for very isolated galaxies, the lensing agrees with MOND predictions up to very large galactocentric distances. Strong lensing in MOND is not expected to deviate much from Newtonian gravity without dark matter \citep{sanders08,tian17, milg20}. 

In the current paper, we investigate in detail PDM halos of point masses and their pairs and the influence of the EFE on them. We also present the predicted maps of gravitational lensing around them. In particular, we do this for the QUMOND modified gravity version of MOND \citep{qumond}. The point masses are assumed to represent galaxies, galaxy groups and clusters as a zeroth order approximation. We investigate whether the special effects of MOND distinguishing it from the \lcdm cosmology are within the reach of the current instruments. The topic is particularly timely since the recently launched \textit{Euclid} satellite is expected to revolutionize the field of weak lensing. It will survey about 1/3 of  the sky with an angular resolution comparable to the \textit{Hubble} Space Telescope. The multitude of the photometric filters will not only allow determining the redshifts of the lensed galaxies, but also of the accurate stellar masses of the lenses, that are needed for predicting the gravitational field in MOND. Unlike the distinguishing test of MOND and particle dark matter based on dynamical measurements, weak lensing does not impose any assumptions on the dynamics of the investigated objects (such as being in virial equilibrium) and is insensitive to many of the uncertain aspects of galaxy evolution.

The paper is organized as follows. We explain how to calculate the PDM density in QUMOND in \sect{mond}. Then we remind the weak lensing formalism in general in \sect{lens}. We progress to the case of a single point mass in \sect{pm}. We discuss the case when there is no external field (\sect{pmzefe}), external field with a known direction (\sect{pmefeknown}) and external field with an unknown direction (\sect{pmefeunknown}). We also compare the QUMOND predicted PDM halo to a dark halo in the \lcdm cosmology in \sect{pmNFW}. Section~\ref{sec:tpm} is devoted to a pair of point masses, observed either perpendicularly to the connecting line (\sect{tpmlens}), or along it (\sect{tpmlenslos}). The external field is treated in an approximate way and, therefore, this approximation is justified in detail in \sect{tpmefe}. In \sect{obs}, we estimate whether the deviation of QUMOND from Newtonian gravity can be achieved with the current instruments and we find that it indeed seems feasible. We summarize our findings in \sect{sum}.  

{In this paper, the symbol $f()$ denotes an unspecified function of the parameters in parenthesis. The natural logarithm is denoted as $\ln$, while the decadic logarithm as $\log_{10}$.}

\section{Calculating the effective distribution of the lensing mass in modified gravity} \label{sec:mond}

We base the present work on the QUMOND theory, since it provides a particularly simple way to calculate the density of PDM \rph. It can be calculated as 
\begin{equation}
\rph = \frac{-1}{4\pi G}\nabla\left[\nu\left(\frac{g_N}{a_0}\right)\mathbf{g_N}\right],
\label{eq:pdm}
\end{equation}
where $\mathbf{g_N}$ stands for the gravitational acceleration calculated in the usual Newtonian way from the distribution of the baryonic matter, $a_0$ stands for the acceleration constant of the MOND theory, $a_0\approx1.2\times 10^{-10}$\,m\,s$^{-2}$ \citep{mcgaugh11,mcgaugh16,lelli17}, and $\nu$ for the interpolating function. The theory does not give a concrete prescription for $\nu$, but it dictates its behavior in two limits: 1) $\nu(x)\approx 1$ for $x\gg 1$, such that the Newtonian gravity is restored for high accelerations, and 2) $\nu(x)\approx x^{-1/2}$ for $x\ll 1$, such that flat rotation curves of galaxies are reproduced at small accelerations. In this paper, we assume the observationally motivated $\nu$ function
\begin{equation}
    \nu(x) = \frac{1}{1-e^{-\sqrt{x}}}
    \label{eq:nu}
\end{equation}
that can explain  the rotation of H\,I gas clouds in galaxies excellently \citep{mcgaugh16,li18}. In this paper, we were interested in gravitational lensing around a point mass or a pair of point masses. For these simple mass configurations, we could obtain analytic expressions for the density of PDM  effectively. For more complex distribution of sources, public codes exist that can evaluate the density distribution of PDM numerically \citep{por,raymond}.

In \app{rhosig} we derive general functional forms for the density and surface density of the PDM in the deep-MOND regime of any MOND theory in which the concept of PDM makes sense and does not introduce any other new constants of physics than $a_0$. Namely, if $M$ denotes the total baryonic mass of the system, $d$ its characteristic size, $g_e$ the magnitude of the external field, and $\vec{\alpha}$ a vector of dimensionless parameters that describe the geometry of the problem, then:
\begin{equation}
\rph(\vec{x}) = \frac{1}{d^2}\sqrt{\frac{Ma_0}{G}}f\left( \vec{\alpha}, \frac{\vec{x}}{d}, \frac{g_e d}{\sqrt{GMa_0}}\right)-\rho_\mathrm{b}(\vec{x}),
\label{eq:gendens}
\end{equation}
and
\begin{equation}
\sigph(\vec{\xi}) = \frac{1}{d}\sqrt{\frac{Ma_0}{G}}f\left( \vec{\alpha}, \frac{\vec{\xi}}{d}, \frac{g_e d}{\sqrt{GMa_0}}\right)-\Sigma_\mathrm{b}(\vec{\xi}),
\label{eq:sigphdmgen}
\end{equation}
where $\rho_\mathrm{b}$ and $\Sigma_\mathrm{b}$ denote the density and surface density of the baryonic matter, respectively.

\section{Methods to predict the maps of gravitational lensing}\label{sec:lens}
In this section, we review the basics of gravitational lensing. We  follow \citet{bartelmann17} { with some details complemented from \citet{bartelmann01}}.
When the light emitted by  a distant source, typically a galaxy, passes around a massive object, the lens, located between the source and the observer, its path is bent. As the result, the observer captures a deformed image of the source. In the weak lensing limit, that is when the deformations are small, the transformation of the image can be considered linear, such that the images of circular sources are ellipses. Such a linear mapping can be understood as a combination of flattening, rotation, magnification and shift of the original image.

If the linear approximation is not sufficient, we speak about strong lensing. {It was shown for spherically symmetric lenses, that MOND predicts only a small deviation from the Newtonian gravity without dark matter in the strong lensing regime \citep{milg12}. This prediction agrees with observations \citep{tian17}. Later in this paper we confirm that MOND does not give rise to strong lensing in its weak-field regime even for lenses without the circular symmetry. }

Most observational characteristics of gravitational lensing can be derived from the lensing potential $\psi$. If the thickness of the mass distribution of the lens is negligible compared to the distances to of the lens from the source and from the observer, which is the case of the astrophysical situations of interest of this paper, then we can make the so-called thin lens approximation. Then the lensing potential can be calculated as
\begin{equation}
    \Delta\psi(\vec{\theta}) = 2\frac{\Sigma(\vec{\theta})}{\sigcr},
    \label{eq:poiss}
\end{equation}
where
\begin{equation}
    \Sigma(\vec{\theta}) = \int_{-\infty}^\infty \rho(\vec{\theta},z) \mathrm{d}z,
    \label{eq:sd}
\end{equation}
and
\begin{equation}
    \sigcr^{-1} = \frac{4\pi G}{c^2}\frac{D_L D_{LS}}{D_S}.
    \label{eq:sigcr}
\end{equation}
In these formulas ${\vec\theta} = (\theta_1, \theta_2)$ is a vector, expressed in the units of radians, in a Cartesian system that touches the celestial sphere at the position of the lens from the point of view of the observer. For circularly symmetric lenses we match the origin of the coordinate system with the center of the symmetry.  The symbol $\Delta$ stands for the two-dimensional Laplace operator with respect to the coordinates \vec{\theta}. The surface density $\Sigma$ in \equ{sd} is given by integration of the volume density $\rho$ of the lens along the line of sight $z$. In the context of this paper, $\rho$ is the sum of the density of the baryonic matter and the PDM. The constant \sigcr is called the critical surface density.  In \equ{sigcr}, $D_L$ denotes the {angular-diameter} distance of the lens from the observer, $D_S$ the {angular-diameter}  distance of the source from the observer, and  $D_{LS}$ the {angular-diameter distance of the source for an observer at the lens at the cosmic time when the photon passes the lens.} For a point mass $M$ (without a PDM halo) we have
\begin{equation}
    \psi(\vec{\theta}) = \frac{4GM}{c^2}\frac{D_{LS}}{D_LD_S}\ln\theta.
    \label{eq:pointlens}
\end{equation}

The Jacobian matrix characterizes locally the deformation of the image by the lens. It is defined as
\begin{equation}
    A_{ij}(\vec{\theta}) = \delta_{ij}-\frac{\partial^2\psi(\vec{\theta})}{\partial \theta_i\partial \theta_j},
    \label{eq:jacobi}
\end{equation}
where $\delta_{ij}$ is Kronecker's symbol. If we calculate the eigenvectors
$\vec{e_+}$ and $\vec{e_-}$ and the corresponding eigenvalues 
$\lambda_+$ and $\lambda_-$ of the inverse of the Jacobian matrix, $A^{-1}$, we learn to what ellipse a circular source is deformed into.  The major axis of the ellipse points in the direction of $\vec{e_+}$ that corresponds to the larger eigenvalue $\lambda_+$. The ratio of the eigenvalues is equal to the ratio of the axes of the ellipse. The {complex ellipticity (one of the two common definitions which is related more directly to observations, see \citealp{bartelmann01}) is defined as
\begin{equation}
    \epsilon = \frac{\lambda_+-\lambda_-}{\lambda_+ + \lambda_-} \, e^{2\phi i}, 
\end{equation}
where $\phi$ denotes the position angle of the major axis of the ellipse. We can also express $\epsilon$ as a sum of its real and imaginary part, $\epsilon = \epsilon_1+i\epsilon_2$. We call here by ellipticity (without the complement of ``complex'') the magnitude of the complex ellipticity. } 

We define the shear as 
\begin{equation}
    \gamma = \gamma_1+ i\gamma_2,
\end{equation}
where
\begin{equation}
    \gamma_1 =\frac{1}{2}\left( \frac{\partial^2\psi}{\partial \theta_1^2}-\frac{\partial^2\psi}{\partial \theta_2^2}\right), \quad \gamma_2 = \frac{\partial^2\psi}{\partial \theta_1\partial \theta_2},
    \label{eq:shear}
\end{equation}
{In the case of small deformations of the image, i.e. in the weak lensing limit, 
\begin{equation}
   \epsilon\approx \gamma.
\end{equation}
}

In the lowest order of approximation, the intrinsic {complex ellipticity of the background source, $\epsilon_\mathrm{S}$, and the complex ellipticity} caused by lensing add up, such that the total observed ellipticity of the source is $\epsilon_\mathrm{S}+\epsilon$. This fact, along with the linearity of \equ{poiss} and  \equ{shear}, ensures that the surface density obtained by stacking of many lenses is the average of the surface densities of the individual lenses.

If the surface density of the lens is circularly symmetric, a simpler way of calculating the image deformation exists. { The lens then deforms the image of the source only be stretching or squeezing along circles centered on the lens. We thus introduce the tangential shear as
\begin{equation}
    \gamma_t = -\cos(2\phi)\gamma_1 -\sin(2\phi)\gamma_2.
\end{equation} 
It satisfies $|\epsilon|\approx|\gamma_t|$ in the weak lensing limit.
Let us denote by $R$ the projected radius form the center of the lens (in kpc),  $M(R)$ the projected mass contained within the radius $R$, $\vartheta = R/D_\mathrm{L}$, and  $m(\vartheta) = M(\vartheta D_L)/(\pi D_\mathrm{L}^2 \sigcr)$. Then 
\begin{equation}
    \gamma_t(\vartheta) = \frac{m(\vartheta)}{\vartheta^2}-\frac{1}{2\vartheta}\frac{\mathrm{d}m(\vartheta)}{\mathrm{d}\vartheta}.
    \label{eq:shearsym}
\end{equation}
}

Next, the effects of gravitational lensing can be characterized by magnification. If we define the convergence 
\begin{equation}
 \kappa = \Sigma/\Sigma_\mathrm{crit},
\end{equation}
then magnification 
\begin{equation}
\mu \approx 1+2\kappa.
\label{eq:mu}
\end{equation}
 For axially symmetric lenses observed along the symmetry axes we have
\begin{equation}
    \kappa = -\frac{1}{2\vartheta}\frac{\mathrm{d}m(\vartheta)}{\mathrm{d}\vartheta}.
\end{equation}
{However, for the reconstruction of the density field, the magnification is not as useful as the shear and this is why we focus here on the shear.}

\subsection{Scaling with distance}
\label{sec:distscaling}
{Suppose that we have already calculated the lensing potential and shear for some distance of the lens, $D_{L,0}$, source, $D_{S,0}$ and their mutual distance, $D_{LS,0}$. It would be useful to know how these results will change if the same lens is at $D_L= a D_{L,0}$, and the source at $D_S= b D_{S,0}$. Because angular diameter distances do not add up linearly, the distance between the lens and source will change to $D_{LS} = c D_{LS,0}$, where $c$ depends on $D_L$ and $D_S$. From \equ{pointlens} and the linearity of \equ{poiss}, we obtain that the new lensing potential $\psi(\theta)  = \frac{c}{ab}\psi_0(a\theta)$, where $\psi_0$ is function prescribing the original lensing potential.  From here and \equ{shear} we get that  the new shear $\gamma$ is related to the original shear $\gamma_0$ as $\gamma(\theta) = \frac{ac}{b}\gamma_0(a\theta)$. Therefore, only the lensing ellipticity changes, not the positional angle. }

{ \section{Distances of the lenses and sources in the considered models}
\label{sec:modeldist}
In all models considered in this paper, the lens is assumed to lie at the distance of $D_L = 200\,$Mpc and the sources  ``in infinity''. We opted for this value of $D_L$ because it is a rather typical   value for the largest existing catalog of isolated galaxy pairs \citep{nottale18} and the lensing by such objects is the primary objective of this paper. The future catalogs of isolated galaxy pairs will probably consist of relatively nearby galaxies too, because the measurements of radial velocities are necessary. At the distance of $D_L = 200\,$Mpc, one arcsecond almost exactly corresponds to one kiloparsec. 

We checked that the ratio $D_{LS}/D_S$ appearing in \equ{sigcr} deviates from unity by less than four percent for redshifts higher than two and by less than six percent for redshifts higher than one. The numbers assume the final Planck cosmology \citep{planck13}. For the purpose of this exploratory paper, we assume $D_{LS}/D_S = 1$. The shear published here can be rescaled to other distances of the sources by multiplying  by the corresponding ratio $D_{LS}/D_S$.}

\section{Gravitational lensing by a point mass in QUMOND}
\label{sec:pm}

\subsection{Zero external field}
\label{sec:pmzefe}
Let us review the case of point mass that is in a zero external field. Gravitational acceleration $\vec{g}$ of any spherically symmetric object in QUMOND can be calculated as \citep{qumond}:
\begin{equation}
    \vec{g} = \vec{g_N}\nu\left(g_N/a_0\right).
    \label{eq:mond}
\end{equation}
By applying the Poisson equation to this field, one gets the density of the PDM. From \equ{mond}, one can see that $\vec{g} \approx \vec{g_N}$ in the region where $g_N\gg a_0$, that is well below the so-called MOND transitional radius 
\begin{equation}
 \rM = \sqrt{\frac{GM}{a_0}}.
 \label{eq:rm}
\end{equation}
Therefore, in the vicinity of the point mass the PDM halo nearly has a zero density if the interpolation function reaches one  sufficiently rapidly for large arguments \citep{milg08}. This is the case of the interpolation function \equ{nu} assumed here. Well beyond the transitional radius, $g \approx \sqrt{g_Na_0}$. Then, for an isolated point mass, we obtain that the PDM density is equivalent to that of an isothermal sphere \citep[e.g.,][]{milg13} 
\begin{equation}
\rho_\mathrm{IS} = \frac{1}{4\pi r^2} \sqrt{\frac{Ma_0}{G}},
\label{eq:rhoIS}
\end{equation}
which has the surface density of 
\begin{equation}
\Sigma_\mathrm{IS} = \frac{1}{4r}\sqrt{\frac{Ma_0}{G}}. 
\label{eq:sigIS}
\end{equation}
\citet{milg08} explored the surface distribution of PDM for a variety of circularly symmetric sources and interpolation functions. They found that MOND predicts rings in the surface density of PDM, unless the interpolation function reaches the high-acceleration limit too slowly. The maximum surface density of the ring is on the order of the constant $\Sigma_0 = a_0/G = 8.6\times10^{14}\,M_\sun\,$Mpc$^{-2}$ and it is reached approximately at the distance of $\rM$.

\begin{figure}
        \resizebox{\hsize}{!}{\includegraphics{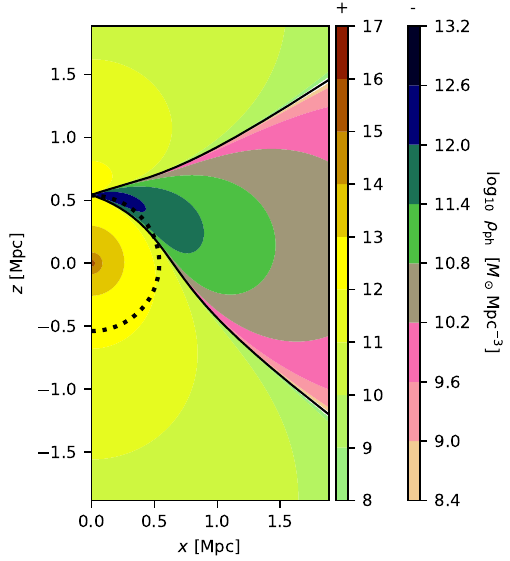}}
        \caption{Distribution of the PDM density around a point mass in a homogeneous external gravitational field. The mass of the point source, located in the center of the coordinate system, is $10^{11}$\,M$_\sun$. The external field has the value of $0.02\,a_0$ and points up. The left and right color scales indicate the positive and negative densities of PDM, respectively. The dotted half-circle marks the theoretical estimate of the radius $\req$ below which the effects of the external field are negligible. } 
        \label{fig:pmefe}
\end{figure}

\begin{figure}
        \resizebox{\hsize}{!}{\includegraphics{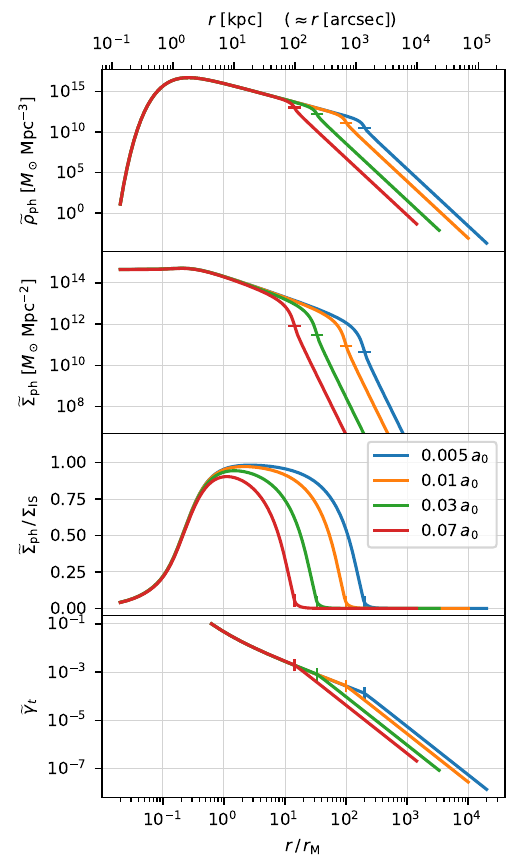}}
        \caption{Stacked point masses in randomly oriented external fields.  From top to bottom: {\bf Row 1}: Density of PDM.   {\bf Row 2}: Surface densities obtained by the projection of the volume densities above. {\bf Row 3}: Ratio of the surface density of the PDM halo and the surface density of the approximating isothermal sphere. 
        {\bf Row 4}: Tangential shear.  The different lines correspond to the indicated various intensities of the external field. The small horizontal line on each curve marks the corresponding radius \req. For the given models, 1\,kpc corresponds to 1.03\arcsec.} 
        \label{fig:pmeferhosig}
\end{figure}

\subsection{Known direction of the external field}
\label{sec:pmefeknown}
Each object in the real Universe is however embedded in a nonzero external gravitation field that comes from the neighboring galaxies or the large-scale structures (\sect{intro}). The external field is expected to affect the PDM halos at galactocentric distances larger than about the radius
\begin{equation}
\req = \frac{\sqrt{GMa_0}}{g_e}, 
\label{eq:req}
\end{equation} where the gravitational acceleration generated by the point mass $\sqrt{GMa_0}/r$ equals the external gravitational acceleration $g_e$ \citep{milg83a,milgmondlaws}. In our exploration of the PDM density distribution for a point mass in an homogeneous external field, we made use of \equ{pdm}. We assumed that the point mass is in the centre of a Cartesian coordinate system and the external gravitational field points in the direction of the positive $z$-axis. The resulting expression for the PDM density is stated in \app{pm}. In \fig{pmefe}, we plotted this density distribution for a point mass $M = 10^{11}$\,M$_\sun$ and an external field $g_e = 0.02\,a_0$. The dotted half-circle shows the radius \req.  One can note from here that the PDM halo is approximately spherical only below $\req$, where the external gravitational field is negligible compared to the field from the point mass.  The central depression of the density of PDM is not visible in this figure because its size is just about 0.01\,Mpc.  Beyond the \req radius, the distribution of PDM has only the cylindrical symmetry. \citet{qumond} derived an analytic expression for the density of PDM of QUMOND far from the point mass. It is approximately proportional to
\begin{equation}
    \rph\propto \frac{1}{r^3}\left(\frac{3z^2}{r^2}-1\right).
    \label{eq:asym}
\end{equation}
The region of the positive dark matter thus has asymptotically a bi-conical shape (see also \citealp{milg86}).

The direction of the external field is relatively easy to determine only for galaxies close to their massive neighbors or galaxy clusters \citep[e.g.,][]{haghi16,muller19,chae20}. Gravitational lensing in these situations will be investigated in detail in \sect{tpm}.

\begin{figure}
        \resizebox{\hsize}{!}{\includegraphics{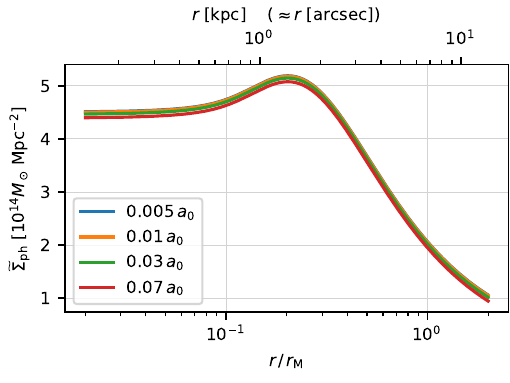}}
        \caption{Detail of the central profile of the surface density of PDM. } 
        \label{fig:sigonly}
\end{figure}

\subsection{Unknown direction of the external field}
\label{sec:pmefeunknown}
 In this section, we aim to predict and explore the density profiles of  PDM halos that would be inferred by a weak lensing analysis of stacked lenses, supposing that for each lens the external field has a random direction drawn from an isotropic distribution. As we explained in \sect{lens}, the surface density we obtain in this way is the average surface density over all lenses. The average surface density $\sigphtilde$ can be obtained by projecting the spherically averaged volume density 
\begin{equation}
    \rphtilde(r) = \frac{1}{4\pi r^2} \int_{\mathbb{S}(r)}\rph(\vec{r^\prime}) \mathrm{d}S.
\end{equation}

We calculated the radial profiles of $\rphtilde$ and $\sigphtilde$ numerically for a point mass of the mass of $4\times10^{10}\,M_\sun$. We varied the magnitude of the external field between $0.005-0.07a_0$. The results are plotted in the first two panels of Fig.~\ref{fig:pmeferhosig}.  We also plotted the surface density of PDM in another scale in \fig{sigonly}, in order to better illustrate the ring-of-dark-matter phenomenon.  The vertical dashed lines in these figures indicate the radius \rM. The third panel of \fig{pmeferhosig} shows the ratio of  $\sigphtilde$ to the surface density of an isothermal halo  according to \equ{sigIS}. The fourth panel shows the {tangential shear}. {  These results were calculated using \equ{shearsym}, assuming the distances from  \sect{modeldist}}. The radii \req are marked in \fig{pmeferhosig} by the short horizontal bars on the corresponding curves. One can note several facts in these plots:

\begin{enumerate}
    \item The profile of the density or surface density of the PDM halo is strongly affected by the external field only in the regions that are further from the point mass than \req. We need to measure surface density with a precision better than about $10^{12}\,M_\sun\,$Mpc$^{-2}$. The differences in the \req radii are in the order of hundreds of kpc for the range of external field magnitudes considered here.  
    \item Between \rM and \req, the external field reduces the surface density by about one to ten times compared to the isolated case described by \equ{sigIS}.
    \item Under the \rM radius, the external field does not affect the surface density substantially.
    \item The spherically averaged density reaches a maximum at 0.26\rM, the surface density at 0.20\rM, with a change at third valid decimal place when the external field intensity ranges between (0.005-0.07)$a_0$. The position of the maximum can however be different for other interpolating functions, see \citet{milg08}.
    \item The ring-of-dark-matter phenomenon occurs, for the assumed interpolation function, near the radius at which $\epsilon  = 1$, that is where strong lensing occurs and the weak lensing approximation is no longer precise.
    \item The spherically  averaged density of the PDM halo  $\rphtilde$ falls steeply beyond \req. This can already be recognized from  \equ{asym} for the approximate asymptotic density of a PDM halo, since the integral of this expression over a sphere with a radius $r$ is zero. We found numerically that   
\begin{equation}
 \rphtilde\propto r^{-7}, \quad   \sigphtilde\propto r^{-6}
 \label{eq:densefe}
\end{equation}
for $r\gg \req$.

One can make a similar calculation to that in \app{rhosig} to derive a general equation for the density of the PDM  in the deep-MOND regime of any MOND theory which does not introduce any other new constant than $a_0$.  The result is:
\begin{equation}
\rphtilde(r) = \frac{1}{r^2}\sqrt{\frac{Ma_0}{G}}\,f\left(\frac{g_e r} {\sqrt{GMa_0}}\right).   
\label{eq:densefegen}
\end{equation}
For QUMOND, a comparison to \equ{densefe} yields that $f(x)\propto x^{-5}$ for $r \gg \req$, and thus:
\begin{equation}
\rphtilde \propto \frac{1}{r^7}\frac{G^2M^3a_0^3}{g_e^5}.  
\label{eq:densefequmond}
\end{equation}
in this region. We found numerically that the proportionality constant is close to $(4\pi)^{-2}$.

\end{enumerate}

Equation~\ref{eq:densefegen} tells us how to rescale these results for points of different masses, as long as we are interested in the regions in the deep-MOND regime. For example, if we are interested in an object that is four times more massive than the fiducial one, we have to look for a curve for an external field that if two times stronger and than to multiply the values in the plot by two. Alternatively, we can have a look at a radius that is two times larger and divide the value which we read in the plot by four.

\begin{figure}
        \resizebox{\hsize}{!}{\includegraphics{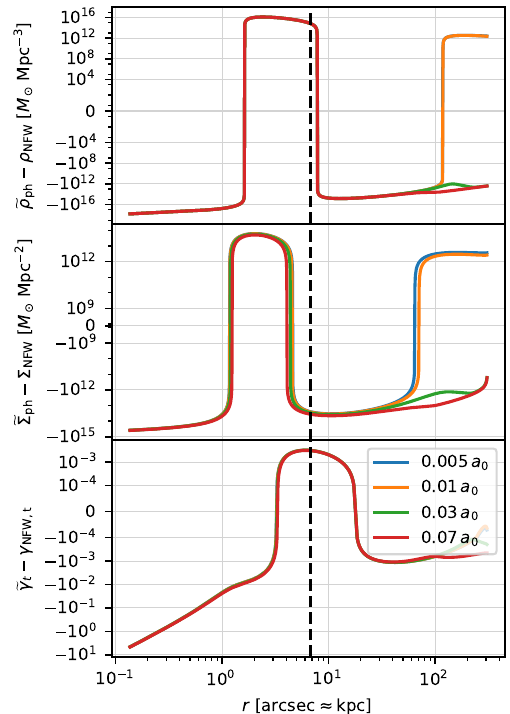}}
        \caption{Comparison of a PDM halo to an NFW halo. The point mass has a baryonic mass of $4\times 10^{10}\,M_\sun$ and lies at the distance of 200\,Mpc.  \textbf{Row 1}: Difference of densities (the curves for $g_e = 0.005\,a_0$ and $0.01\,a_0$ nearly coincide). \textbf{Row 2}: Difference of surface densities.  \textbf{Row 3}: Difference of {tangential shears}. The sources are assumed to be at infinity. The vertical dashed line indicates the MOND transitional radius \rM. } 
        \label{fig:nfw}
\end{figure}

\subsection{Comparison to an NFW halo}
\label{sec:pmNFW}
Between the radii \rM and \req, the density of PDM is roughly like of an isothermal sphere, given by \equ{rhoIS}. This is in contrast with the predictions of the \lcdm cosmology, where  dark matter halos are expected to have a density profile similar to the Navarro-Frenk-White (NFW) profile \citep{nfw}.  We thus could discriminate between the \lcdm model and MOND if we could decide what profile of (phantom) dark halos  galaxies have. We  quantified the difference for one representative galaxy. We chose the baryonic mass of the galaxy as $M = 4\times 10^{10}\,M_\sun$, similar to the characteristic mass $M_*$ of the galaxy mass function \citep{kawinwanichakij20}.  Then we assigned a dark matter halo to the galaxy making use of the mean stellar-to-halo relation \citep{behroozi13} and  halo mass-concentration relation \citep{diemer15} at zero redshift. This yielded the virial mass of the halo of $1.6\times10^{12}\,M_\sun$ and a virial radius of 310\,kpc.  We studied the radial range only up to the virial radius (see, e.g., \citealp{diemer15} for a more detailed behaviour of \lcdm halos at large radii). 

We compared this NFW halo with the spherically averaged MOND PDM halos in \fig{nfw} for several different magnitudes of the external field: the top panel shows the difference of the profiles of density, and the middle and bottom panel the same for surface density and {tangential shear}, respectively.  For the calculation of the lensing effects, we used \equ{shearsym} {and the distances from \sect{modeldist}}. 

The figure shows that for detecting the deviation of the PDM halo from the NFW halo need to detect a difference in surface densities on the order of $10^{13}\,M_\sun\,$Mpc$^{-2}$ or in {tangential shears} on the order of $10^{-3}$ in the radial range of tens of arcsec. We will discuss whether such a measurement is feasible in \sect{obs}. The deviation of the PDM halo from the NFW halo beyond about 100\,kpc depends strongly on the magnitude of the external field.

\section{Gravitational lensing by two point masses in MOND}
\label{sec:tpm}

\citet{milg86} showed that there is a region of negative PDM between a pair of two masses near the point of zero acceleration. For the case of equal point masses, he derived, assuming the AQUAL version of MOND, an expression for the density of PDM in the plane going through the zero-acceleration point and is perpendicular to the line joining the point masses. He found that \rph reaches an  infinitely negative density near the symmetry axis since $\rph\propto r^{-1/2}$. The region of the negative PDM has a shape of the figure ``8'' rotating by its short axis around the line connecting the two point masses.   In  \sect{tpmgen}, we explore the  distribution of PDM in more detail, assuming the QUMOND formulation of MOND. The weak lensing signature of QUMOND  for this mass configuration is studied in \sect{tpmlens}.

\begin{figure*}
        \centering
        \includegraphics[width=17cm]{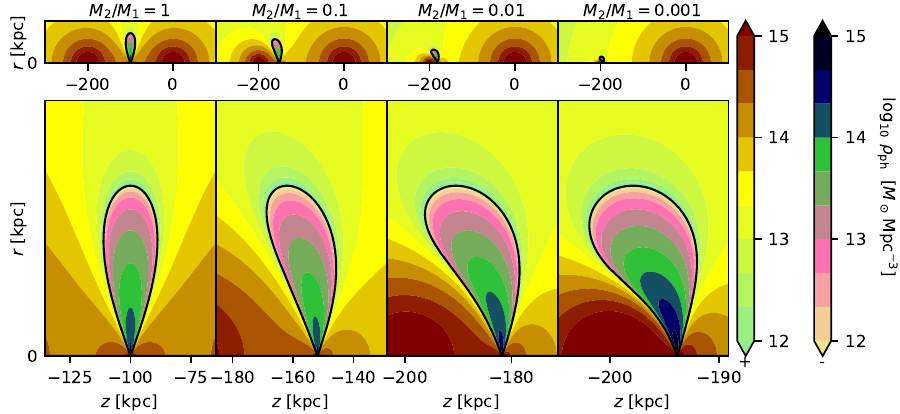}
        \caption{Distribution of the PDM around two point masses (zero external field assumed) with various mass ratios indicated above each column. The mass of $M_1$ is $10^{11}\,M_\sun$ and the two point masses are 200\,kpc apart. The mass ratio is the same for each column. The top row shows the whole system, in the bottom row we can see magnified the region of the negative PDM. Note that the vertical ranges of the plots in the bottom row are different for each plot. In each plot, the $r$ and $z$ axes have the same scale.}
        \label{fig:tpm}
\end{figure*}

\begin{figure}
        \resizebox{\hsize}{!}{\includegraphics{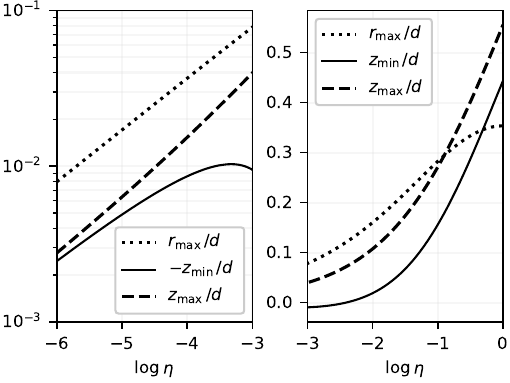}}
        \caption{Dimensions of the region of negative PDM.}
        \label{fig:negregsize}
\end{figure}

\begin{figure}
        \resizebox{\hsize}{!}{\includegraphics{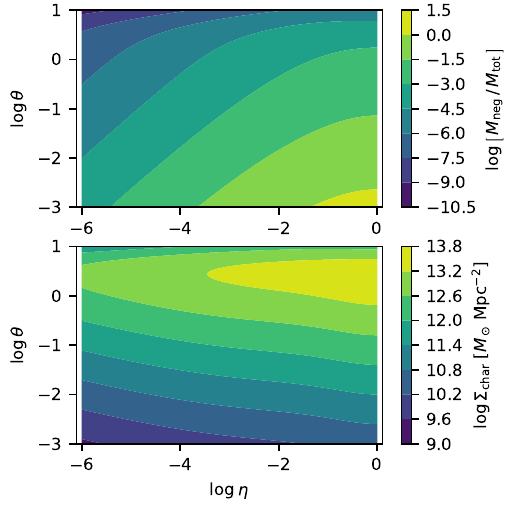}}
        \caption{Top: Mass of the region of the negative density  of PDM in the units of the total baryonic mass of the system $M_\mathrm{tot} = M_1+M_2$ as a function of the parameters $\eta = M_1/M_2$ and $\theta = \sqrt{Ga_0M_\mathrm{tot}}/d$. Bottom: Characteristic projected surface density of the region of the negative PDM. }
        \label{fig:mnegrat}
\end{figure}

\subsection{Distribution of PDM around two  point masses}
\label{sec:tpmgen}
In this section, we will investigate the distribution of PDM around a pair of point masses in a zero external field. Let us denote the masses of the two points by $M_1$ and $M_2$, such that $M_1\geq M_2$, and their distance by $d$. We will work in the cylindrical coordinate system that has its axial coordinate, $z$ coincident with the connecting line of the point masses. We denote the radial coordinate by $r$. Let the point mass $M_1$ be located at $z=0$ and $M_2$ at $z=-d$. The density \rph  or projected surface density \sigph of the PDM at the position $\vec{x}$ can depend only on $G, a_0, M_1, M_2, d$ and $\vec{x}$. For dimensional reasons,  \rph and \sigph must be given by expressions of the forms: 
\begin{equation}
\begin{split}
\label{eq:rhosig}
    & \rph = \frac{\mtot}{d^3} f\left(\eta, \theta, \frac{\vec{x}}{d} \right)\quad \textrm{and} \\
    & \sigph = \frac{a_0}{G} f\left(\eta, \theta, \frac{\vec{x}}{d} \right),
\end{split}
\end{equation}
where 
\begin{equation}
    \mtot = M_1+M_2, ~~~ \eta = \frac{M_2}{M_1}, ~~ \textrm{and} ~~ \theta = \frac{\sqrt{G\mtot/a_0}}{d}.
\end{equation}
The parameter $\theta$ quantifies how deep in the MOND regime the system is. In the following, we will consider only situations where $\log_{10} \theta >-3$ because, for lower values, a real system would probably be affected by the external field from the large-scale cosmic structures (\sect{intro}). This is expected to happen approximately for $\theta<g_e/a_0$, as we precise in \sect{tpmefe}. If we deal with regions of space where the gravitational acceleration is much less than $a_0$, then we can directly use Eqs.~\ref{eq:gendens} and ~\ref{eq:sigphdmgen}, with $d$ signifying the distance between the point masses.

We derived the analytic expression for the density of PDM around two point masses without an external field, making use of \equ{pdm}. It is stated in Appendix~\ref{app:tpm}. Let us first explore the shape of the region of the negative PDM density. In \fig{tpm} we plotted the PDM for two point masses that are in the distance of 200\,kpc and various mass ratios $\eta$ for $M_1 = 10^{11}\,M_\sun$. The top row of the figure, showing the whole system, illustrates that with decreasing $\eta$ the negative PDM region becomes smaller and gets closer to  $M_2$. In the bottom row, the displayed region scales with the size of the region of the negative PDM. The images here show that the negative PDM region becomes bent towards the mass $M_2$ with decreasing $\eta$ and the typical density of the region of negative PDM becomes more negative. Equation~\ref{eq:rhosig} allows  the maps of \rph to be rescaled for different masses and separations of the point masses. 

The characteristic signature of MOND is the region of the negative PDM. For its detection by weak lensing it is useful to know its size.  We now prove a noteworthy fact that {as long as the interpolation function $\nu$ is monotonically decreasing,}  the shape of the region of negative PDM in QUMOND is given by a fully Newtonian expression, that is that the expression does not contain the MOND constant $a_0$ or the interpolation function. This actually holds true for any configuration of matter, as long as we are concerned with the regions where the density of the real matter is zero. Indeed, using Einstein's notation, \equ{pdm} gives outside the distribution of the real matter that 
{\begin{equation}
    -4 \pi G \rph  = \partial_i\left[ \nu g_{N}^i\right] = \partial_i(\nu)g_N^i+\nu\partial_ig_N^i  = \frac{\nu^\prime}{a_0}\partial_i(g_N^jg_{N,j})^{1/2}g_N^i.
\label{eq:zerorhoph}
\end{equation}
In the last expression, the first factor can only be negative, and therefore the surface where $\rph=0$ is given by the other two Newtonian factors. }

We characterize the shape of the negative PDM region by its maximum radial extent \rmax and its minimum and maximum $z$ coordinate with respect to the position of the mass $M_2$ (i.e., $\zmin=0$ indicates that the minimum $z$ coordinate is reached at $z=-d$). It follows from the previous paragraph that any linear dimension of this region, $x$, depends only on $G, M_1, M_2$, and $d$. A~dimensional analysis then indicates that $x/d = f\left(\eta\right)$. We derived the function numerically for \rmax, \zmin and \zmax and is shown in  \fig{negregsize}. The value of \zmin becomes negative for $\eta<3.91\times 10^{-3}$.

Next, let us focus on the mass and characteristic surface density of the region of negative PDM.  A~dimensional analysis gives that the mass of the region, \mneg, satisfies $\mneg/\mtot = f(\eta, \theta)$. We calculated the function numerically in the top panel of \fig{mnegrat}. It is interesting that the mass of the negative region can be comparable, or even higher than the total  baryonic mass of the system. This happens preferably for systems consisting of two similar masses in deep-MOND regime. For example, for two galaxy clusters with a mass of $10^{13}\,M_\sun$ located $15\,$Mpc apart, the $\theta$ parameter is about $10^{-2}$ and therefore the region of negative PDM would have a mass of a galaxy cluster.

The surface density depends on the particular configuration of the problem and orientation with respect to the line of sight. We explore some particular situations in detail in \sect{tpmlens}. Here we study only the characteristic surface density of the region of the negative PDM to get an order-of-magnitude estimate and investigate how it scales with $\eta$ and $\theta$. In particular, we  defined $\Sigma_\mathrm{char} = \mneg/\left[ 2\rmax (\zmax-\zmin)\right]$. It is plotted in the bottom panel of \fig{mnegrat} as a function of $\eta$ and $\theta$. The figure indicates that the characteristic surface density depends primarily on $\theta$ and that it reaches a maximum at $\eta = 1$ and $\theta \approx 2.5$, i.e. for equal-mass systems near the border of the Newtonian and deep-MOND regime. This is similar to the case of the surface density of the PDM around a point mass, which reaches the maximum  near the border of the Newtonian and deep-MOND regions. For example, for two equal galaxies with a mass of $10^{11}\,M_\sun$ (comparable to the Andromeda galaxy), $\theta = 2.5$ corresponds to a distance of 6\,kpc. Such galaxies would be interacting. Figure~\ref{fig:negregsize} shows that the radius of such a region would be about 3\,kpc. 

Next, we explored the limit behavior of \rph close to the zero-acceleration point located at $(r,z) = (0,z_0)$  where $z_0 = -d/(1+\sqrt\eta)$.  After calculating the second-order Taylor expansion of the expression for the Newtonian gravitational potential around the zero-acceleration point, and making use of the fact that $\nu(x)\sim x^{-1/2}$ for small arguments, we got from  \equ{pdm} that
\begin{equation}
\begin{split}
    & \rph \sim \frac{\left(\!\!\sqrt{M_1}+\sqrt{M_2}\right)^2}{8\pi d^{3/2}(M_1M_2)^{1/4}}\sqrt{\frac{a_0}{G}}\frac{8(z-z_0)^2-r^2}{\left[4(z-z_0)^2+r^2\right]^{5/4}} \\ & \textrm{for}~~(r,z)\rightarrow(0,z_0).
    \label{eq:sing}
\end{split}
\end{equation}
Thus, if we approach the zero-acceleration point perpendicularly to the $z$ axis, $\rph$ reaches a negative infinity like $r^{-1/2}$. The same was shown for the AQUAL formulation of MOND by \citet{milg86}. On the other hand, when we approach the zero-acceleration point along the line $r=0$,  \rph reaches the positive infinity like $ (z-z_0)^{-1/2}$. Interestingly, this holds true even if the point masses are close to each other, such that the system is in the Newtonian regime.

While the volume density of PDM is diverging around the zero acceleration point, in \app{finite}, we prove that the projected surface density near this point is finite. So far, it has only been proven that there is an upper limit on the surface density of PDM halos of nearly spherical objects \citep{milgmondlaws}.

\subsection{Pair of point masses observed perpendicularly}
\label{sec:tpmlens}
Lensing analyses are based on observations ellipticity, position angles and magnifications of the lensed sources. A map of these quantities can be converted to the map of surface density (e.g., \citealp{kaiser93}). In this section, we provide these maps for a pair of point masses predicted by MOND. In this first investigation of the topic, we consider just two orientations of the point masses with respect to the observer: the perpendicular view and the axial view.  Furthermore, we consider lensing by what we call the superposition density. This is the density which is obtained by assigning the point masses the spherically averaged PDM densities $\rphtilde$ derived in \sect{pm}, that are expected by for single point masses in the respective external field, and summing them. The difference of PDM density and the superposition density quantifies a key property of MOND, namely its non-linearity, by which it differs from the Newtonian gravity. In addition, we propose to test MOND by subtracting the weak lensing signal observed around solitary point masses from the lensing signal observed around pairs of point masses and to compare the result to the MOND predictions predictions. A similar approach is used for example for detecting the lensing signal of the cosmic filaments connecting galaxy groups \citep{epps17,yang20}. In such a test, it would be necessary to ensure that the solitary point masses and pairs of point masses reside in equally strong external fields. {In \sect{obs}, we discuss the prospects of detecting the specific signatures of QUMOND observationally.}

\subsubsection{Computational methods}
\label{sec:comp}
Let us start with the case that the point masses are observed perpendicularly to their connecting line.  We calculated the surface density map simply by integrating numerically the expression for the PDM of a pair of point masses in Appendix~\ref{app:tpm}: 
\begin{equation}
\sigph = \int_{-Y}^{Y}\rph(x,y,z)\mathrm{d}y.  
\label{eq:tpmsig}
\end{equation}
Note that the expression for \rph does not involve the external field. In the following, we took into account the value of the external field $g_e$ approximatively  through the limit of the integral, $Y$. In particular, we integrated between the outermost points where the Newtonian acceleration generated by the point masses was equal to the Newtonian external field,  $g_{e,N}$, a number obtained by solving \equ{mond} for $g = g_e$. In other words, we took into account only the  PDM located in the region where the total gravitational field is dominated by the contribution of the point-mass pair. Consequently, \sigph was considered non-zero only in a limited region of space.  This approximation is justified in \sect{tpmefe}. It is related to the finding from  \sect{pm} that the density of PDM drops quickly outside the region dominated by the internal field. We refrained in this first investigation of the topic the averaging \rph over all possible direction of the external field, like in the case of a solitary point mass in \sect{pm}, because of high computing demands. 

In order to calculate lensing maps, shown in Figures~\ref{fig:mapsel}-\ref{fig:mapspadiff}, we had to solve the two-dimensional Poisson equation \equ{poiss} for the lensing potential. This was done by numerically by convolution of $\sigph$ with the lensing potential of a point mass given by \equ{pointlens} and adding the contributions of the real matter. To make the computation feasible, we progressed in a zoom-out fashion. First, the potential was calculated on a fine grid in the region depicted in the maps. Then it was necessary to add also the contributions to the lensing potential coming from the outside of  this region. In the second step, we thus defined a grid that was twice as large as the previous one, had the same number of grid nodes, and had the same center. The nodes that were covered already by the previous grid were assigned a zero surface density and the others the correct density \sigph. The lensing potential at this grid was again calculated by convolution. We repeated this procedure until we covered the region of non-zero \sigph  completely.  We added also to the surface density the contributions of the point masses themselves once they got covered by the expanding grid. In this way, the contributions to the lensing potential from various spatial scales were contained in several layers.  At the end, the lensing potential in each layers was interpolated for the grid points of the densest grid and the resulting lensing potential was obtained by summing up the contributions from all layers. The correct functioning of the code was verified by comparison with analytic solutions for the lensing potential for various combinations of point masses and isothermal spheres. Numerical derivatives were used to obtain the Jacobian matrix from the lensing potential according to \equ{jacobi}.

For the axial configuration, i.e. when the two point masses lie along the same line of sight, we obtained the surface density of the PDM in a similar fashion:
\begin{equation}
\sigph = \int_{Z_1}^{Z_2}\rph(x,y,z)\mathrm{d}z.  
\end{equation}
The limits of the integral were again chosen as the minimum and maximum $z$-coordinate at which, for a given $x$ and $y$, the Newtonian acceleration generated by the two point masses was equal to  $g_{e,N}$. Given that this density distribution is axially symmetric, we could use the simple \equ{shearsym} to obtain the {tangential shear}. Some aspects of gravitational lensing for this configuration of matter have already been investigated by \citet{milg08}. Unlike we do here, they used an approximate formula for the density of PDM and focused on the influence of the choice of the MOND interpolation function.

\subsubsection{Explored sample of models}
\label{sec:models}
The models that we explored were variations of a fiducial model. It is fully characterized by the dimensionless parameters $\theta =0.1$, $\eta=0.4$ and $g_{ext} = 0.03a_0$ and the mass of the more massive point of $M_1 = 10^{11}\,M_\sun$. This implies the mass of the lighter point $M_2 = 4\times10^{10}\,M_\sun$ and the distance between the points of $d = 127.4$\,kpc. {We assume the distances of the lenses and sources as described in \sect{modeldist}. This results in the angular separation of the two point masses of 2.18\arcmin.}

The other explored models were obtained by varying of one of the parameters $\eta$, $\theta$ or $g_{ext}$ of the fiducial model while keeping the other parameters intact. In particular, the explored values were $\eta =[0.1, 0.16, 0.4, 1]$, $\theta = [0.06, 0.1, 0.3, 1]$, $ g_{ext}/a_0 = [0.005, 0.01, 0.03, 0.07]$. The separation of the points of varied as $[212.4, 127.4, 42.5, 12.7]$\,kpc when varying $\theta$ and as $[113.0, 116.0, 127.4, 152.3]$\,kpc when varying $\eta$. At the assumed lens distance, one kpc is almost exactly one arcsecond and the critical density comes out $\log_{10} \sigcr/(M_\sun\,\mathrm{Mpc}^{-2})=15.92$, which can be used for converting the surface density into the lensing magnification with the aid of \equ{mu}. One can rescale the results to other configurations of the problem using the relations \equ{gendens} and~\ref{eq:sigphdmgen} and the results of \sect{distscaling}.

\subsubsection{Results}
\label{sec:tpmres}
\textbf{Surface density.} Figures~\ref{fig:sigmawide} and~\ref{fig:sigmazoom} show the surface densities of PDM around the considered models from the perpendicular view. The former figure shows for all models the whole halos, while the latter shows the vicinity of the point masses. The plots showing the fiducial model are marked by the red frames. The sizes of the displayed regions in the zoomed-in plots are proportional to the distances between the point masses by a constant factor. The size of the PDM halo in the fiducial model, beyond which the halo is expected to be truncated, is around 500\,kpc. This size only mildly depends on the choice of $\eta$ or $\theta$, but decreases strongly with increasing $g_{ext}$. The outline remains in all cases nearly circular, except in the model with the strongest $g_{ext}$, when the halo becomes elongated along the line connecting the point masses. All parameters influence noticeably the inner morphology of the halo, as evident from \fig{sigmazoom}. 

The region of the negative PDM density has its counterpart in as a dip in the the surface density.  Nevertheless, the surface density is not necessarily negative -- the contribution of negative density near the zero acceleration point is balanced by the contribution of the positive density from the regions which are further away from the symmetry axis. The surface density can be negative if the system is exposed to a strong-enough external field. This is because the external field reduces the contribution of the positive PDM do the surface density that otherwise occupies the external regions of the system. The surface density becomes negative also for very low acceleration systems (i.e., the $\theta$ parameter is low). In this case, the positive contribution of the outer part of the PDM halo, whose mass is given by the masses $M_1$ and $M_2$ and the value of the external field, is overweighted by the large amounts of negative PDM near the zero acceleration point. Indeed, if the distance between the two point masses is large, then $\theta$ is small and \fig{mnegrat} indicates that the mass of the region of the negative PDM is large.  

The magnitude of the dip is generally a factor of a few. For the fiducial model, the dip occurs at the surface density level of about $10^{13}\,M_\sun\,$Mpc$^{-2}$. This surface density grows the strongest with the $\theta$ parameter. This is because the density of PDM is the highest near the MOND transitional radius\footnote{The size of the MOND transitional radius is negligible in the images.} (\sect{pm})  and the points become closer to each other with decreasing $\theta$.
We note that even if the PDM density diverges near the zero acceleration point (\equ{sing}), the surface density does not, as  proved analytically in  \app{finite}. 

Another characteristic feature which we see in \fig{sigmazoom} is the offset outer halo of the less massive point. This is best visible in the top-left panel, showing a pair of a high mass ratio. One might look for such an affect for galaxies close to galaxy clusters. 

\textbf{Surface density difference.} Figures~\ref{fig:sigmadiffwide} and~\ref{fig:sigmadiffzoom} show the difference between the PDM surface density and the superposition surface density, in order to explore the effect of the non-linearity of MOND. The first figure is again the complete view of the whole PDM halo, while the second shows only the central region around the point masses. The difference is negative near the center of the halo, meaning that the superposition surface density is higher than the true PDM surface density. This is because the amount of PDM around one of the point masses, compared to the case that the point mass were isolated,  is reduced by the EFE imposed the other point mass. We note that the reduction of the surface density between the two point masses is the opposite of what we expect if the galaxies were connected by a cosmic filament \citep{yang20}.

On the other hand, the superposition surface density is less than the true PDM surface density in the outer parts of the halo. This is caused by the external field in which the point masses are embedded both: the combined PDM halo of the pair of point masses is truncated by the EFE at a larger radius, than the PDM halos of the single point masses that form the superposition halo.   We  note that in the external parts of the halo, the difference is relatively constant, having the value of around $10^{12}\,M_\sun$Mpc$^{-2}$. 

{In the theories postulating particle dark matter and Newtonian gravity, the real density and superposition density will differ too, but probably in a different way. During the first approach of the two halos toward each other, the tidal force will bulge the halos along their connection line, producing a surplus of dark matter in between of the galaxies. Later, once the halos experience tidal stripping and compactification, the result is more difficult to predict. Nevertheless,  \citep{pawlowski17b} inspected the distribution of satellites around isolated pairs of galaxies in a \lcdm simulation and found that they tend to be lopsided toward each other. In contrast, \fig{sigmazoom} shows that the PDM halos in QUMOND  appear ``squashed'' in the region between the point masses. Here, the density of PDM is reduced compared to
the superposition density, see \fig{sigmadiffzoom}. At least this is true for the surface densities. Figure~\ref{fig:tpm} shows that the volume density contours of PDM halos have more intricate shapes.} 

\begin{figure}
        \resizebox{\hsize}{!}{\includegraphics{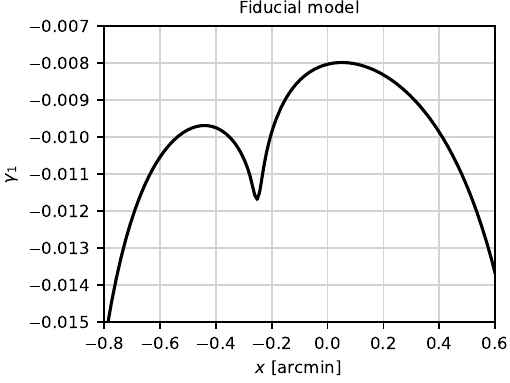}}
        \caption{Profile of {the real part of the shear} along the line connecting two point masses in the fiducial model. {The imaginary part is negligible over the plotted range.} }
        \label{fig:axcut}
\end{figure}

{\textbf{Lensing shear.} Figures.~\ref{fig:mapsgam1} and~\ref{fig:mapsgam2} show the maps of the real and imaginary part of the shear, respectively.  The influence of the changes of the different parameters of the models is theoretically understood easier in the terms of the ellipticity magnitude and position angle. We will do that in the following paragraphs. However, the shear components have the advantage over the ellipticity magnitude and position angle that they average  when stacking  different lenses.}

\textbf{Lensing ellipticity.} Figure~\ref{fig:mapsel} presents the maps of lensing ellipticity for our models. For each model, there are three maxima of ellipticity: two near each of the point masses, and one near the zero-acceleration point. Detection of this last peak would be a smoking gun evidence for modified gravity. This peak is absent in Newtonian gravity for spherical DM halos, as we detail below. The first column of the figure demonstrates that the position the zero-acceleration peak depends on the mass ratio of the two point masses. The magnitude of the peak does not seem to depend significantly on the mass ratio of the point masses for the range of models considered here.  Figure~\ref{fig:axcut} depicts for the fiducial model the profile of the {real component of the shear} along the line connecting the two point masses.  {Because the imaginary part is much smaller than the real part, the plotted quantity, in absolute value, corresponds to the lensing ellipticity.} It shows that for detecting the feature, the measurement of ellipticity has to reach the precision of $10^{-3}$ on the angular scale of 0.1\arcmin. 

The second column of \fig{mapsel} shows that  the lensing ellipticity increases as the two point masses are becoming closer to each other. This can be understood in the following way. Equation~\ref{eq:sigphdmgen} tells us that surface density of the PDM is proportional to $d^{-1}$ as long as we deal with a system in a deep-MOND regime, which means, in the present situation, as long as the separation between the point masses is much greater than the sum of their MOND transitional radii. Then, the linearity of Eqs.~\ref{eq:poiss} and~\ref{eq:shear}  implies that lensing ellipticity is proportional to $d^{-1}$ too. The last column of \fig{mapsel} indicates that the lensing ellipticity in the vicinity of the point masses varies only very little for the range of the strengths of the external field considered in this study. This is because the surface density of PDM plus baryonic matter, which is determines the lensing effects  (\equ{poiss}), depends most on the inner, that is the densest part of the PDM halo. The external field removes only the outer, little dense part of the PDM halo.

\textbf{Lensing position angle.} Let us turn to \fig{mapspa} which shows the maps of the position angle of the lensing shear ellipses for our models. The position angle of 0 or 180\degr correspond to the orientation along the horizontal direction. The figure shows that the shear ellipses near the zero acceleration point are always elongated perpendicularly to the line connecting the point masses. In spite of the divergent and discontinuous nature of the PDM density near the zero acceleration point (\equ{sing}), the lensing effects are finite and continuous. It is striking that the maps of positional angle are virtually identical for all models in the second and third column. In the second column the distance between the point masses in kpc or arcmin varies, but displayed portion of the sky is proportional to the distance of the point masses. As before, we can use Eqs.~\ref{eq:sigphdmgen}, \ref{eq:poiss}, and~\ref{eq:shear} to explain why the maps of shear positional angle do not depend on the separation of the point masses. The separation affects the lensing potential only as a multiplicative factor and thus only the magnitude of the complex shear is affected (i.e. the ellipticity), but not its phase (i.e. the positional angle). As before, the EFE only affects the low-density regions of the PDM distributions, that are contribute only little to the surface density of the PDM, and therefore the maps of positional angle do not vary noticeably in our models when varying the intensity of the external field in the third column of \fig{mapspa}.

{\textbf{Lensing shear difference.} 
A key property of MOND is that the gravitational fields of two objects do not add up.  In Figs.~\ref{fig:mapsgam1diff} and~\ref{fig:mapsgam2diff} we show the difference between actual shear expected by QUMOND and that stemming from the superposition density. The difference is on the order between about $10^{-4}$ and $10^{-2}$, both for the real and imaginary components. Again, we aim to understand the behavior of with variations of the different parameters theoretically in the following paragraphs in the terms of lensing ellipticities and position angles.}

\textbf{Lensing ellipticity difference.} {Figure~\ref{fig:mapseldiff} shows the differences in ellipticity caused by the PDM density and ellipticity caused by the superposition density.  The difference is the highest when the point masses are the closest to each other.} These maps emphasize the peak of ellipticity near the zero acceleration point, since this peak is absent in the superposition model. The peak in these ellipticity difference maps is on the order of $10^{-2}$. The spatial size of the peak is roughly proportional to the distance between the point masses.  The EFE, at least within the range considered in this work, has again no appreciable impact on the result.

\textbf{Lensing position angle difference.} The differences between the shear positional angle caused by the PDM density and the superposition density  are displayed in \fig{mapspadiff}. In most regions, the difference of the position angles is a few degrees.  Near the zero acceleration point, the positional angle difference is close to zero.  The misalignment that is the easiest to detect observationally is around ninety degrees. There indeed are such regions.
A comparison to \fig{mapsel} and \fig{mapseldiff} however reveals that this happens only where both that the actual and superposition ellipticities are close to zero, which is unfortunately where it is the most difficult to measure the positional angle. In order to simplify the comparison with \fig{mapsel}  we drew in the upper halves of \fig{mapspadiff} the contours of constant lensing ellipticity from  \fig{mapsel} as the dotted lines. The contours in are at the ellipticity levels of $10^{-2}$, $10^{-2.5}$ and $10^{-3}$. The deviation of a few tens of degrees is encountered at the ellipticities of a bit above $10^{-3}$. For close pairs (high $\theta$), the situation is more positive, because the ellipticites can be around $10^{-2}$, but only at in a very small region of space.

Unlike for the positional angle, the difference of the positional angles of the true and superposition PDM shows a dependence on the distance between the point masses. This is because the superposition PDM is a construct that is inconsistent with MOND, such that \equ{sigphdmgen} does not hold true for it, because the superposition density does not obey the space-time scaling symmetry.

\begin{figure*}
        \centering
        \includegraphics[width=17cm]{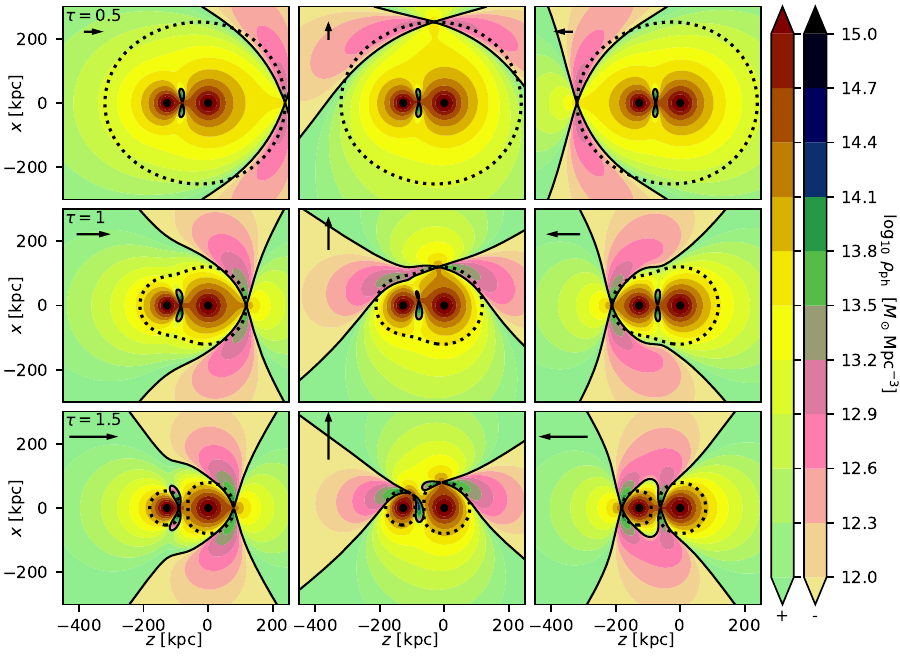}
        \caption{Pair of point masses in an external field. The image shows the density of the PDM in a plane going through the point masses. Different columns correspond to different directions of the external field. The direction is indicated by the arrows. In each column, the magnitude of the external field is stronger with each tile downwards. The magnitude is specified through the $\tau$ parameter defined by \equ{tau}. The dashed curves indicate the surfaces within which we neglected the EFE in \sect{tpmlens}. The full black lines indicate the boundaries between the regions of positive and negative density of the PDM.  }
        \label{fig:tpmefewide}
\end{figure*}

\begin{figure*}
        \centering
        \includegraphics[width=17cm]{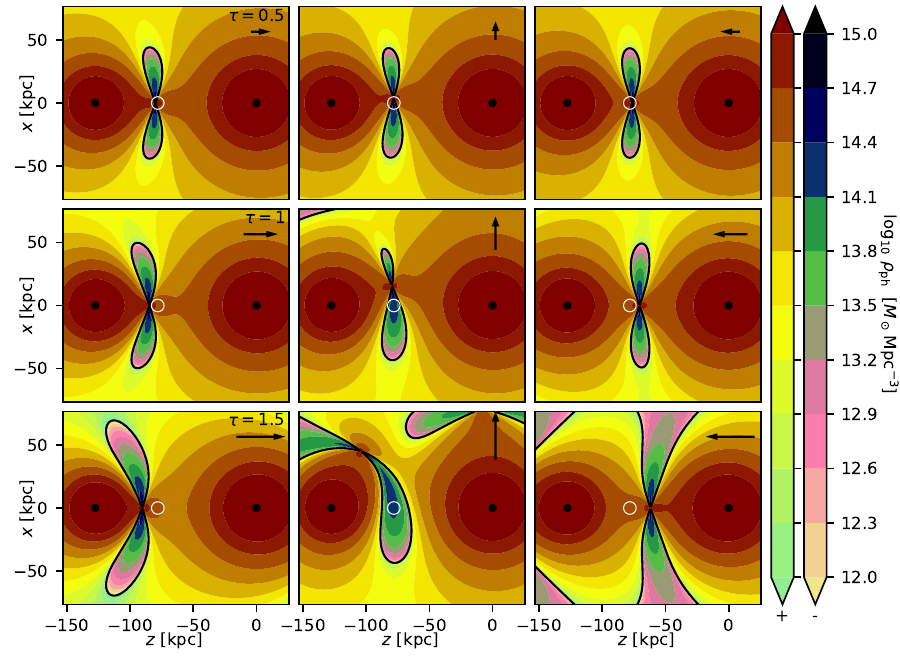}
        \caption{As in \fig{tpmefewide}, but only the near vicinity of the point masses is shown. The white circle indicates the position of the zero-acceleration point in the situation without an external field. The radius of the circle is 5\,kpc.}
        \label{fig:tpmefenarrow}
\end{figure*}

\begin{figure*}
        \centering
        \includegraphics[width=17cm]{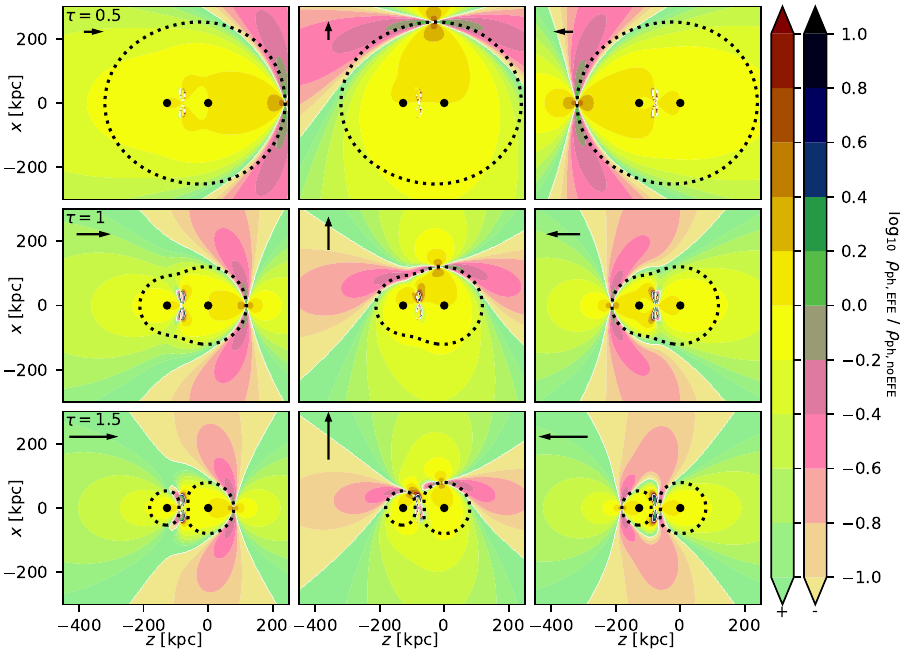}
        \caption{Ratio of the densities of PDM around two point masses without and with the external field. The meaning of the different tiles is the same as in  \fig{tpmefewide}.  }
        \label{fig:tpmeferatio}
\end{figure*}

\subsection{Pair of point masses observed aligned}
\label{sec:tpmlenslos}
Let us now turn to the case of two point masses observed along their connecting line, elaborating the results of \citet{milg08}.   The projected surface density for this orientation is axially symmetric, and therefore we will investigate only the radial profiles of surface density and {tangential shear}.  We explored the same set of models as for the perpendicular view, see \sect{models}. 

\textbf{Surface density.} The results for surface density are presented in Figs.~\ref{fig:lossig} and~\ref{fig:losell}. The profile shows, similarly to the case of a single point mass, a peak near the MOND transitional radius, forming a ring when observed on the sky \citep{milg08}. The surface density of the PDM near the ring is about $10^{15}\,M_\sun\,$Mpc$^{-2}$. The height of the peak grows  with the $\eta$ parameter, since the total mass of the system increases. Increasing the $\theta$ parameter (i.e. decreasing the separation between the point masses while keeping their masses fixed), leads to a decrease of the surface density at a given distance from the origin\footnote{It might be confusing that \sigph here decreases with increasing $\theta$, while it was increasing for the perpendicular view in the \fig{sigmazoom}, where we showed that this is a universal consequence of \equ{sigphdmgen}. The contradiction is just apparent: in \fig{sigmazoom}, we plotted a spatial range that is fixed in the units of the distance between the points, while the range plotted in \fig{mapspadiff} is fixed in the units of kpc or arcsec. In other words, in \fig{sigmazoom}, with increasing $\theta$, we were plotting smaller and smaller region around the system center, where the density is the highest. If the profiles of \sigph in \fig{mapspadiff} are plotted with the radial distance expressed in the units of $d$, then indeed \sigph increases with $\theta$ at fixed $r/d$. }. This is because the gravitational field of one point mass is progressively more reduced by the EFE coming from the other point mass, as $\theta$ increases, such that the density of the PDM decreases. If the point masses are far from each other, such that they reduce the gravity of each other the least, then \equ{sigIS} tells us that the surface density is proportional to $\sqrt{M_1}+\sqrt{M_2}$. In the opposite limit, when the two point masses coincide, the same equation indicates that the surface density is proportional to $\sqrt{M_1+M_2}$, that is lower. The strength of the external field again  only has a negligible impact on the appearance of the inner profile of the PDM halo.

\textbf{Surface density difference.} The difference of the surface densities of PDM and the superposition density is generally on the order $10^{-13}\,M_\sun\,$Mpc$^{-2}$. Surprisingly, this is rather universal value valid for all radial distances from the symmetry center and for most of the explored models. The shapes of the profiles beyond about $10^0-10^1$\,kpc are caused by the same effects as discussed for the perpendicular configuration. Near the center of symmetry, the superposition surface density can be both higher or lower than the superposition surface density. To see why, we note in Fig.~\ref{fig:pmeferhosig} for a single point mass, that the density of the PDM around the point mass is falls steeply toward zero. The surface density near the point mass is therefore determined by the density further away from the point mass. If two point masses seen along the line-of-sight are close to each other (i.e., low $\theta$), the external field of one of them reduces the volume density of the PDM of the other point mass. This phenomenon is absent for the PDM density and therefore its surface density can be higher than that of the actual PDM if the point masses are close to each other. On the other hand, if the point masses are far apart, the regions close to each of the point masses are almost equal to the superposition densities. Nevertheless, the combined gravitational fields of the two point masses resist easier to the global external field than if the point masses were isolated. Thus, in this case, the superposition PDM surface density is lower than the actual PDM surface density. The enhanced resistance of the actual PDM density to the global external field, when compared to the superposition PDM density, explains the behavior for the varying intensity of the global external field.

\textbf{Tangential shear.} In \fig{losell}, we can see that the radial profiles of {tangential shear} are rather featureless. They are approximated well by a power law with a slope close to 1.5, a value in between of what we expect for a point mass and an isothermal PDM halo, see  Eqs.~\ref{eq:shearsym} and~\ref{eq:sigIS}. The {tangential shear} increases at a fixed radius with increasing $\eta$, because the total baryonic mass of the system increases. With increasing $\theta$, the {tangential shear} decreases as the PDM density of one point mass is becomes diminished by the EFE coming from the other point, which is located closer. The global external field manifests itself virtually only as a truncation of the PDM halo at larger projected galactocentric radii. 

\textbf{{Tangential shear} difference.}  The differences in the true {tangential shear} from the {tangential shear} expected from the superposition density is on the order of $10^{-4}$. The explanation of the behavior of the curves when $\theta$ and the global external fields are varied is the same as for the case of the surface density.

\subsection{Influence of external field}
\label{sec:tpmefe}

In this section, we briefly explore the influence of an external field on the distribution of PDM around a pair of point masses. We also discuss the correctness of the simplified treatment of the external field in   \sect{tpmlens} on the prediction of gravitational lensing. We work in a Cartesian coordinate system whose $z$ axis is defined as before (\sect{tpmgen}) and the $x$ and $y$ axes are chosen such that the coordinate system is right-handed. We derived an analytic expression for the density of PDM. It is stated in \app{tpmefe}. It allows for an arbitrary orientation of the external field.

For the purpose of the discussion of the correctness of \sect{tpmlens}, we introduce a parameter that characterizes how important the external field is for the density of the PDM between the two point masses. 
We expect that the distribution of the PDM will be relatively unaffected in the region  where the intensity of the internal field of the system will be weaker than the intensity of the external flied, $g_e$. The size of this region, $r$, thus approximately satisfies $\sqrt{GM_\mathrm{tot}a_0}/r\approx g_e$. We expect that the region between the point masses starts being affected by the external field once $r$ becomes smaller than $d$, which is equivalent to $\theta<g_e/a_0$. This leads us to the introduction of the  $\tau$ parameter --
\begin{equation}
    g_e/a_0 = \tau\, \theta
    \label{eq:tau}
\end{equation}
-- which is expected to be much less than one if the impact of the external field on the distribution of PDM between the two point masses is negligible, and much greater than one when the impact of the external field is substantial. We verify this below.

In \fig{tpmefewide}, we show the variations of the density of PDM in our fiducial model (\sect{models}) when the system is embedded in a homogeneous external field of different directions and intensities. The intensities are quantified through the parameter $\tau$. In the depicted  models, $\tau=1$ corresponds to the external field of $g_e = 0.1 a_0$. Figure~\ref{fig:tpmefenarrow} shows the same but the displayed region is zoomed in to the vicinity of the two point masses. In this figure, in addition there is a white circle of the radius of 5\,kpc that is centered on the zero-acceleration point of the fiducial model without the external field. The dotted lines indicate the borders of the region of isolation defined as in \sect{comp}, where we used it to determine the limits of the integral for calculation of the surface density of the PDM. Figure~\ref{fig:tpmeferatio} shows the ratio of the models with external field to the model without the external field from \sect{tpmres}. We can make several observations in these figures.

Figures~\ref{fig:tpmefewide} and~\ref{fig:tpmefenarrow} show that the density of PDM is only little affected by the direction of the external field in the region of isolation.  Figure~\ref{fig:tpmefenarrow} shows that this density in this region is almost the same as if the external field was neglected. This justifies the integrand of \equ{tpmsig}. The limits of the integral in \equ{tpmsig} are justified by our results from \sect{pm} that after averaging the PDM densities over all possible directions of the external field, the density steeply decreases beyond the outer border of the region of isolation. 

In contrast,  just beyond the region of isolation the density of PDM depends strongly on the intensity and direction of the external field. It is noteworthy that the external region of  negative PDM density always nearly touches the region of isolation.  Far from the point masses, the distribution of PDM approaches that of a single point mass in an external field (\equ{asym}).

It is interesting to note that the external field does not much modify the absolute value of the density of the PDM outside of the region of isolation; rather only its sign. The EFE  thus seems to arise from a large part because of the  influence of the regions of the negative PDM density, that mitigate the influence of the regions of the positive PDM density. This is a hypothesis that needs a more careful analysis in future works. When averaging over various directions of the external field, the contributions of the positive and negative densities of PDM almost cancel each other, which explains the steep decline of the curves in \fig{pmeferhosig}  beyond the \req radius. This is hinted also by \equ{asym}, whose spherical average is zero at any radius.

The figures reveal that the parameter $\tau$ indeed  characterizes the importance of the external field on the distribution of the PDM between the pair of point masses, in agreement with the above order-of-magnitude estimates. We verified that this is the case also for models with different values of $\theta$ and $\eta$. At $\tau=1.5$, the density of PDM between the two point masses is strongly affected by the external field. It is difficult to predict what would be the average PDM density in this region when averaging over all possible directions of the external field. This is the reason why we avoided strong external fields in \sect{tpmlens} or very separated point masses in the predictions for gravitational lensing, given that we used an approximate method. The greatest value of $\tau$ used in our lensing models (\app{var}) is 0.7.

The external gravitational field should be dominant also around the point of zero internal acceleration. From this perspective, it is interesting that the ``8'' shaped region of negative PDM near the zero acceleration point is not much influenced by the external field as long as $\tau\lesssim 1$. The position and size of the region vary just by several percent. In weak lensing counterpart of this region will then be just mildly blurred by the varying direction of the external field, as long as the $\tau<1$.

\section{Observational aspects}
\label{sec:obs}

\begin{figure}
        \resizebox{\hsize}{!}{\includegraphics{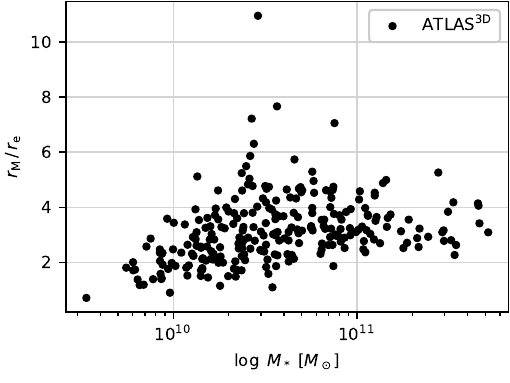}}
        \caption{Comparison of the MOND transitional  radii \rM to the effective radii \reff of the early-type galaxies from the ATLAS$^\mathrm{3D}$ survey. } 
        \label{fig:a3d}
\end{figure}

Here we comment on the observational aspects of the detection of the signatures of QUMOND gravity in weak lensing analysis. First, in MOND, for certain interpolation functions, PDM halos can have central cavities (\sect{pm}). This might have already been observed for one galaxy cluster \citep{jee07,milg08}. It would be desirable to look for similar rings of dark matter also in other objects. Let us turn to galaxies as the lenses. If the effects of the central cavity of the PDM halo are visible only close to the centers of the galaxies, the body of the galaxy might make the determination of the shapes of the background galaxies difficult and therefore practically disable the test. Here we focus on early-type galaxies, because the situation might be much more difficult for the spirals, which contain a lot of substructure. In \fig{a3d} we compared the effective radii of early-type galaxies from the \atlas sample \citep{cappellari11a} to their MOND transitional radii \rM. We estimated the baryonic masses of the galaxies as the dynamical masses within  1\reff published in \citet{cappellari13a} multiplied by 0.87, to account for  the small contribution of the ``dark matter'' to the dynamical mass \citep{cappellari13a}. Figure~\ref{fig:a3d} shows that the MOND transitional radius is typically  three times larger than effective radius. Let us assume, as a rough estimate, that we can detect background sources which are {projected} further away from the galaxy center than the effective radius (after subtracting its model, probably). Figure~\ref{fig:sigonly} indicates that we would not be able to detect the ``ring of dark matter'' effect. In this regard, quiescent galaxies at high redshifts are more promising, because at a fixed stellar mass, they are a few times smaller than their local counterparts \citep[e.g.,][]{trujillo06,trujillo11,damjanov11}. Nevertheless, one might also inspect the deviation of the observed density profile from the isothermal sphere.  The third and fourth panel of \fig{pmeferhosig} indicate that the central deviation might be observable even for galaxies meeting the local mass-size relation.

In \fig{nfw}, we inspected the deviation of the QUMOND-predicted profile of the PDM from the profile expected in the \lcdm cosmology for NFW halos. Actually, it has already been shown recently that the halos of isolated galaxies prefer MOND over NFW halos \citep{mistele23}. It would be interesting to inspect whether the ratio of the observed surface density to the surface density of the MOND isothermal PDM halo (\equ{sigIS}) shows the curvature because of the EFE, as seen in the fourth panel of \fig{pmeferhosig}.

Next, let us consider the number of lenses that we need in order to detect the specific signatures of QUMOND. \citet{bartelmann17} give several useful numbers and formulas. If we have a spatial bin containing $N_\mathrm{b}$ background galaxies, the uncertainty of the measurement of complex ellipticity {projected to come coordinate axis}, $\delta\epsilon$, can be estimated as $\delta\epsilon = \sigma_\epsilon/\sqrt{N_\mathrm{b}}$ where  $\sigma_\epsilon = 0.2$. The number of the background sources can be expressed as $N_\mathrm{b} = n\, S\! N_\mathrm{L}$ if $n$ denotes the number of the background sources per square {arcminute}, $S$ the angular area of the bin and $N_\mathrm{L}$ the number of the stacked lenses. Combining these two equations, we get 
\begin{equation}
N_\mathrm{L} = \frac{\sigma_\epsilon^2}{n S \delta\epsilon^2}.   \label{eq:nl}
\end{equation}
\citet{bartelmann17} state that $n=30-50$ galaxies per square {arcminute} for images by space-based telescopes. In the examples below, we will assume the optimist value {$n=50\,$arcmin$^{-2}$}. The recently launched \textit{Euclid} satellite is expected to have similar capacities and will cover one third of the extragalactic sky.

In \fig{nfw}, we see that the QUMOND {tangential shear} for the considered model differs from the {tangential shear} in \lcdm in the radial range of ca. 4-20\arcsec. The difference is $10^{-3}$. If we use all background galaxies in such an annular aperture, we need to stack 2\,400 lenses for a one sigma detection of this feature, or nine times as much for a three sigma detection. 

For the fiducial model of the pair of point masses in the perpendicular view, {the $\gamma_1$ dip at} the zero-acceleration point has the size of about $0.1\times0.1\arcmin$ and is {depressed} by about  $10^{-3}$ {below} its neighborhood (Figs.~\ref{fig:axcut} and~\ref{fig:mapsel}). Its one sigma detection thus requires 40\,000 lensing pairs. For comparison, the catalog of isolated galaxy pairs of \citet{nottale18} contains just 13\,000 items. In order to detect this feature, a larger catalog of lenses would have to be constructed. Galaxy cluster pairs might be more suitable.

It would be easier to look for the deviation of the actual {shear} from the {shear} resulting from the superposition density. The latter can be constructed empirically, by investigating the lensing around isolated galaxies, similarly to \citet{brouwer21} or \citet{mistele23}.  {Figures~\ref{fig:mapsgam1diff} and~\ref{fig:mapsgam2diff} show} that the difference is on the order of $10^{-3}$ over the area of about eight square arcminutes for our fiducial model. Only 100 lenses are needed for a one sigma detection. {As explained in \sect{tpmres}, in Newtonian gravity with particle dark matter, the action of close-by dark matter halos on each other is probably qualitatively different from the PDM halos in QUMOND.}

Finally, let us mention some {general} potential caveats of similar studies. When determining the baryonic masses of the lens galaxies, one should take into account also the masses of the circumgalactic media \citep[see, e.g., ][]{mistele23}. It can be substantial even for the Milky-Way type galaxies \citep{gupta12}. The catalogs like that of \citet{nottale18} define the isolation in another way than considered here. Namely, they demand the third nearest galaxy to be some multiple of the projected separation of the pair in question. In the present paper, we instead required a homogeneous and rather weak external field. In the catalog of \citet{nottale18} there are some obvious members of galaxy clusters. The two galaxies in the pair just have similar velocities, that are much different from the velocities of the other nearby cluster members. Pairs of nearby galaxies or galaxy clusters can be connected by a cosmic filament \citep{epps17,yang20}, which could interfere with the sought effects. { Also, it should should be stressed that the numbers of necessary lenses given above reflect only the statistical error stemming from the random orientation of the background galaxies. Actually, more lenses might be needed because of other uncertainty sources, such as determining the masses of the galaxies, their distance from the observer, or in the case of object pairs, the distances with respect to each other that are subject to the projection effect.}

\section{Conclusions}
\label{sec:sum}

We expect different gravitational lensing around galaxies (or their clusters) for different solutions to the missing mass problem. Here we investigated in detail the distribution of the PDM around point masses and their pairs in the QUMOND version of MOND, while having in mind the possibility to test the theory by weak lensing.  The results are as follows.

\subsection{Results for a point mass}
\begin{itemize}
    \item The external parts of a PDM halo of a point mass in an external field are offset with respect to the point mass away from the direction of the external field. This is visible in \fig{sigmazoom}.
    \item We derived an analytic expression for the PDM density for this problem (\app{pm}).
\end{itemize}
We considered the radial profile of a PDM halo of a point mass in an external field averaged over all lines-of-sight. This is the profile that is obtained by stacking of many objects embedded in external gravitational fields of random directions. 
\begin{itemize}
    \item As detailed by \citep{milg13}, between the radii of around $\rM$  (\equ{rm}) and $\req$ (\equ{req}), in the absence of an external field, the PDM halo has approximately an isothermal profile, with its density and surface density given by \equ{rhoIS} and \equ{sigIS}, respectively. We found however an external field can can changed these quantities by several tens of percent (\fig{pmeferhosig}). This is effect which can be sought for observationally.
    \item The PDM halo has a central cavity, i.e. its density drops to zero near the point mass. Below \rM, the halo has a deficit of density  with respect to the isothermal profile. If the interpolation function reaches its high-acceleration limit quickly enough, the projected surface density reaches its maximum  near the MOND transitional radius \rM. Observationally, this appears as a ``ring of dark matter'' \citep{milg08}. For the observationally motivated interpolating function adopted here, the ring lies at $0.2\rM$ (\fig{sigonly}). This internal region is not affected by the external field substantially (\fig{sigonly}). While a ring of dark matter has already been reported at a galaxy cluster \citep{jee07}, we found that it might be worth trying to look for it also around elliptical galaxies, particularly the compact ones which are common at high redshifts \sect{obs}.  The ring lies rather in the region probed by the strong gravitational lensing.
    \item At galacticentric radii larger than \req, the density and surface density show a steep decline with radius (\equ{densefe}, \equ{densefequmond}). 
    \item We explained how to rescale the results from \fig{pmeferhosig} and~\ref{fig:sigonly} to other masses \sect{pm} and lens distances \sect{distscaling}. 
    \item We explored  how the predictions of QUMOND and the \lcdm cosmology differ on one example of a galaxy with the baryonic mass of $4\times 10^{10}\,M_\sun$. In \lcdm, the halo has a NFW profile with parameters determined by the scaling relations with the stellar mass. The difference from the logarithmic QUMOND halo (\fig{nfw}) should be easily  distinguishable by observations  (\sect{obs}).
\end{itemize}

\subsection{Results for a pair of point masses}
\begin{itemize}
    \item We derived analytic expressions for the density of PDM around a pair of point masses in zero (\app{tpm}) and non-zero (\app{tpmefe}) external gravitational field.
    \item In agreement with \citet{milg86}, we found that there is a region of negative PDM near the zero acceleration point in QUMOND. We investigated it in detail in \sect{tpmgen}, supposing a zero external acceleration. We investigated its spatial extent (\fig{tpm}, \fig{negregsize}), and mass and typical surface density (\fig{mnegrat}). Interestingly, for realistic situations the mass of the negative PDM can exceed the baryonic mass of the system. The highest surface densities of the are encountered when the system is on the border of the Newtonian and deep-MOND regime, that is when the separation of the point masses $d = \sqrt{GM_\mathrm{tot}/a_0}$.
    \item We found an approximation of the PDM density near the zero-acceleration point (\equ{sing}), assuming no external field. The PDM density reaches both positive and negative infinity at it. We proved, however, that the surface density is finite for any line-of-sight (\app{finite}). {It does not give rise to strong gravitational lensing.}
    \item Weak lensing analyses work with maps of surface density and lensing shear. {We computed their expected maps in QUMOND for the aligned and perpendicular orientation of the two point masses with respect to the line-of-sight. This was done  for several combinations} of the point mass ratios, separations, and external fields and the results were discussed theoretically  (\sect{tpmlens}, \sect{tpmlenslos}).  We also investigated the differences between the predictions of this correct model from a superposition of the PDM halos of two point masses as they would be if they were alone. In practice this correspond to subtracting the weak lensing signal of isolated objects from the signal of binary objects. All resulting maps are shown in \app{var}.
    \item In the perpendicular orientation, QUMOND predicts a peak of lensing ellipticity near the zero-acceleration point. More calculations are needed to see whether this feature will be observable with \textit{Euclid} for galaxies. On the other hand, detecting whether the combined PDM halo of two point masses is not the sum of the PDM halos of single points should not be very difficult.
    \item {Two PDM halos, when coming closer to each other, lose the PDM mass in the space in between of their centers (\sect{tpmres}). The opposite is expected for particle dark matter halos in Newtonian gravity, even if a dedicated analysis is still to be done.}
    \item We also briefly investigated the impact of the external field on the distribution of PDM for the pair of point masses, because, in the previous parts, this was treated approximately. We found that our approximation was reasonable and derived a formula indicating when it is the case (\equ{tau}).
\end{itemize}

\subsection{General results}
\begin{itemize}
    \item We found (\app{rhosig}) a general expressions for the density and surface density of PDM in the deep-MOND regime of a wide class of MOND theories, namely \equ{gendens} and \equ{sigphdmgen}. They are valid as long as the theory does not involve any other new constant of nature than $a_0$ and it is possible to define the concept of PDM in it.
    \item Under the same assumptions, the PDM halo of a point mass in an external field, when averaged over all lines-of-sight, has the radial profile of density given by \equ{densefegen}.
    \item In QUMOND, in vacuum the shape of the region of the negative PDM density does not depend on the interpolation function and the value of $a_0$(\sect{tpmgen}). 
\end{itemize}

We caution that the results obtained here for QUMOND do not necessarily apply to other MOND theories. For example, \citet{gqumond} discussed a MOND theory in which the PDM density near the zero-acceleration point of a pair of point masses is zero. In some MOND theories, it might not even be possible to define the PDM at all.

\bibliographystyle{aa}
\bibliography{citace}

\begin{appendix}

\section{General form of density and surface density of the PDM in the deep-MOND regime}
\label{app:rhosig}
Here we derive the functional form for the density and surface density of the PDM in the deep-MOND regime. It is valid for any MOND theory where the notion of the PDM is applicable and which does not involve any new constant of nature different than $a_0$ (see, e.g., \citealp{gqumond} for examples of those which do). We will derive those for a system of the mass $M$ and of the characteristic size $d$, which is embedded in the external field with the characteristic magnitude of $g_e$. Let us define the effective density  $\rho_E$ as:
\begin{equation}
    \Delta \phi = 4\pi G \rho_E
\end{equation}
where $\phi$ is the MOND gravitational potential. The effective density is thus the sum of the densities of the baryonic and phantom dark matter. For dimensional reasons, $\rho_E$ has to take the form:
\begin{equation}
    \rho_E(\vec{x}) = \frac{M}{d^3}f\left(\frac{MG}{d^2a_0}, \vec{\alpha}, \frac{\vec{x}}{d}, \frac{g_e}{a_0}\right),
\end{equation}
where $\vec{\alpha}$ is a vector of dimensionless parameters that describe the distribution of baryonic matter in the system and the distribution and orientation of the external field in the absence of the system in question. Let us consider a test particle at the position $\vec{x}$ that moves only under the influence of gravity. It is subject of acceleration
\begin{multline}
    \vec{a} = \int \frac{M}{d^3}f\left(\frac{MG}{d^2a_0}, \vec{\alpha}, \frac{\vec{x'}}{d} \frac{g_e}{a_0}\right) K(\vec{x'}, \vec{x}) \,\mathrm{d}^3\!\vec{x'}, \\ K(\vec{x'}, \vec{x}) = \frac{G(\vec{x}-\vec{x'})}{\left|\vec{x}-\vec{x'}\right|^3}.
    \label{eq:a1}
\end{multline}
We will now use the space-time scaling symmetry of the deep-MOND regime for purely gravitating systems of bodies \citep{milg09}, to learn more about the functional form of $\rho_E$. The space-time scaling symmetry tells us that that the previous equality has to hold true also for a system of bodies on trajectories that are expanded by a factor $\lambda$ in space and time, namely:
\begin{equation}
\frac{\vec{a}(\lambda \vec{x})}{\lambda} =  \int \frac{M}{\lambda^3d^3}f\left(\frac{MG}{\lambda^2d^2a_0}, \vec{\alpha}, \frac{\vec{x'}}{\lambda d}, \frac{g_e}{\lambda a_0}\right) K(\vec{x'}, \lambda\vec{x})\, \mathrm{d}^3\!\vec{x'},
\end{equation}
where we used the fact that the transformation changes the acceleration of the test particle to $\vec{a}/\lambda$, just as the external field to $g_e/\lambda$ (because the bodies which generate the field are now further away). After substituting $\vec{x'}/\lambda = \vec{y'}$, one gets
\begin{equation}
\vec{a} = \int \frac{M}{d^3}\frac{1}{\lambda}f\left(\frac{MG}{\lambda^2d^2a_0}, \vec{\alpha}, \frac{\vec{y'}}{d}, \frac{g_e}{\lambda a_0}\right) K(\vec{y'}, \vec{x})\, \mathrm{d}^3\!\vec{y'},
\end{equation}
which can be rearranged to
\begin{equation}
\vec{a} = \int \frac{M}{d^3}\frac{1}{\lambda}f\left(\frac{MG}{\lambda^2d^2a_0}, \vec{\alpha}, \frac{\vec{y'}}{d}, \frac{g_e d}{\sqrt{GMa_0}}\right) K(\vec{y'}, \vec{x})\, \mathrm{d}^3\!\vec{y'}.
\label{eq:a2}
\end{equation}
The requirement of the space-time scaling symmetry implies that the previous equality cannot depend on the choice of $\lambda$. That is satisfied only if 
\begin{multline}
\rho_E(\vec{x}) =  \frac{M}{d^3}\frac{1}{\lambda} \left(\frac{MG}{\lambda^2d^2a_0}\right)^{-1/2}\!\!f\left(\vec{\alpha}, \frac{\vec{x}}{d}, \frac{g_e d}{\sqrt{GMa_0}}\right) = \\ = \frac{1}{d^2}\sqrt{\frac{Ma_0}{G}}f\left( \vec{\alpha}, \frac{\vec{x}}{d}, \frac{g_e d}{\sqrt{GMa_0}}\right).
\end{multline}
From here we get the expression for the PDM density:
\begin{equation}
\rph(\vec{x}) = \frac{1}{d^2}\sqrt{\frac{Ma_0}{G}}f\left( \vec{\alpha}, \frac{\vec{x}}{d}, \frac{g_e d}{\sqrt{GMa_0}}\right)-\rho\textbf{}(\vec{x}).
\end{equation}
Integrating this along a line-of-sight, we get the functional form of the surface density of the PDM:
\begin{equation}
\sigph(\vec{\xi}) = \frac{1}{d}\sqrt{\frac{Ma_0}{G}}f\left( \vec{\alpha}, \frac{\vec{\xi}}{d}, \frac{g_e d}{\sqrt{GMa_0}}\right)-\Sigma_\mathrm{b}(\vec{\xi}),
\end{equation}
where $\vec{\xi}$ denotes the position in the coordinates in the projection plane.

\section{Surface density of the PDM near the zero-acceleration point is finite}
\label{app:finite}
{Here we prove that the surface density of the PDM near the zero-acceleration point is finite, that is that in its vicinity, $|\sigph(\vec{\xi})|$ is smaller than a certain finite constant for any point in the lens plane $\vec{\xi}$. If $\vec{e}$ denotes a unit vector toward the observer, we have:
\begin{equation}
 |\sigph(\vec{\xi})| \equiv \left|\int_{-\infty}^\infty \rph(\vec{\xi}+\tau \vec{e})\mathrm{d}\tau\right| \leq  \int_{-\infty}^\infty \left|\rph(\vec{\xi}+\tau \vec{e})\right|\mathrm{d}\tau.   
\end{equation}
The surface density can come out to be infinite only due to the diverging density near of the zero acceleration point because: 1) \rph is continuous at other points, 2) in the real word,  \rph drops quickly far enough from the two point masses because of the EFE from the large-scale cosmic structure, see \equ{asym}. To inspect the finiteness of the surface density, it thus is enough to show that the last integral converges only for $|\tau|$ smaller than some given constant $L>0$ in the vicinity of the zero-acceleration point. In this region we can approximate \rph by \equ{sing}. If we use a coordinate system whose $z$ coordinate has the origin in the zero acceleration point, we have:
\begin{equation}
|\rph| \propto   \frac{|8z^2-r^2|}{\left(4z^2+r^2\right)^{5/4}} \leq \frac{4}{\left(4z^2+r^2\right)^{1/4}} \leq \frac{4}{(z^2+r^2)^{1/4}}.
\label{eq:est}
\end{equation}
The first inequality can be proved easily by substituting $z = \frac{1}{2}\zeta \cos\alpha$ and $r=\zeta \sin \alpha$. The last expression of \equ{est} has a spherical symmetry around the origin, so, without loss of generality, it is sufficient to examine the convergence of the integral for an observer located at the $z$ axis. Then we get:
\begin{equation}
    \int_{-L}^{L} |\rph(z,r)|\, {\rm d}z 
    \leq 2\int_{0}^{L} \frac{4\, {\rm d}z}{(z^2+r^2)^{1/4}}  
    \leq 8\int_{0}^{L} \frac{{\rm d}z}{z^{1/2}} 
    \leq 8\sqrt{L}.
\end{equation}
This finishes the proof that the surface density of the PDM surrounding a pair of point masses embedded in an external field is finite.}

\onecolumn
\section{Variations of the fiducial model}
\label{app:var}

\begin{figure*}[b!]
        \centering
        \includegraphics[width=17cm]{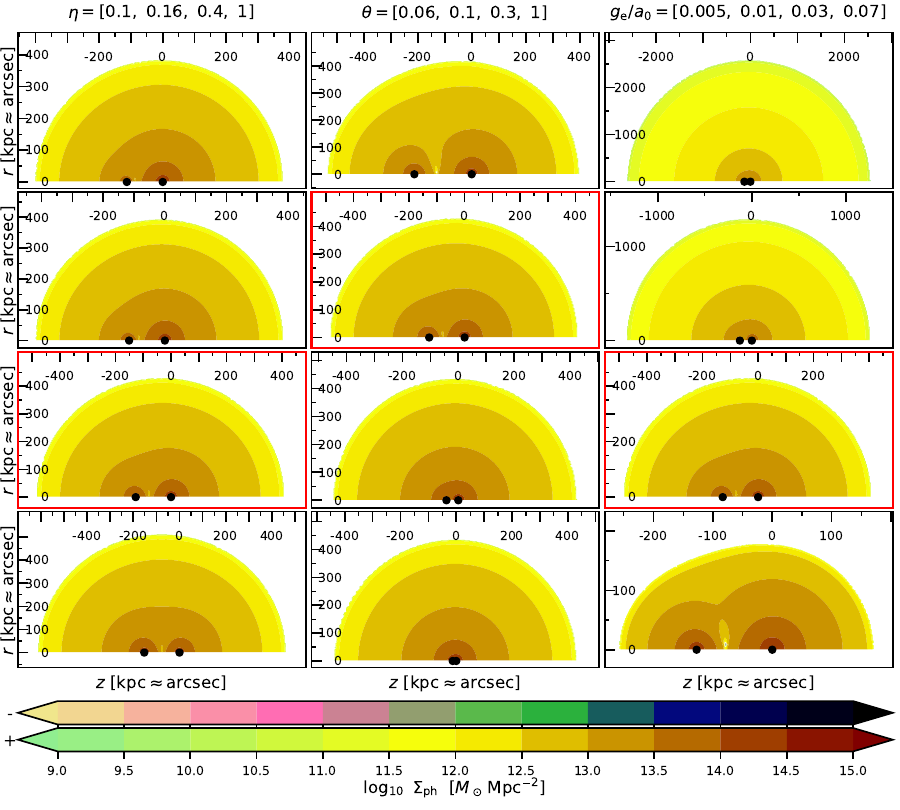}
        \caption{Surface density of PDM around a pair of two point masses. The point masses are seen from a direction perpendicular to their connecting line. The whole halo is shown. In each column, one of the parameters of the fiducial model is varied, as indicated in the titles of the columns. The numbers in the brackets  from left to right, indicate the tiles in the column from top to bottom. The tiles in the red frame show the fiducial model.  }
        \label{fig:sigmawide}
\end{figure*}
\clearpage

\begin{figure*}[b!]
        \centering
        \includegraphics[width=17cm]{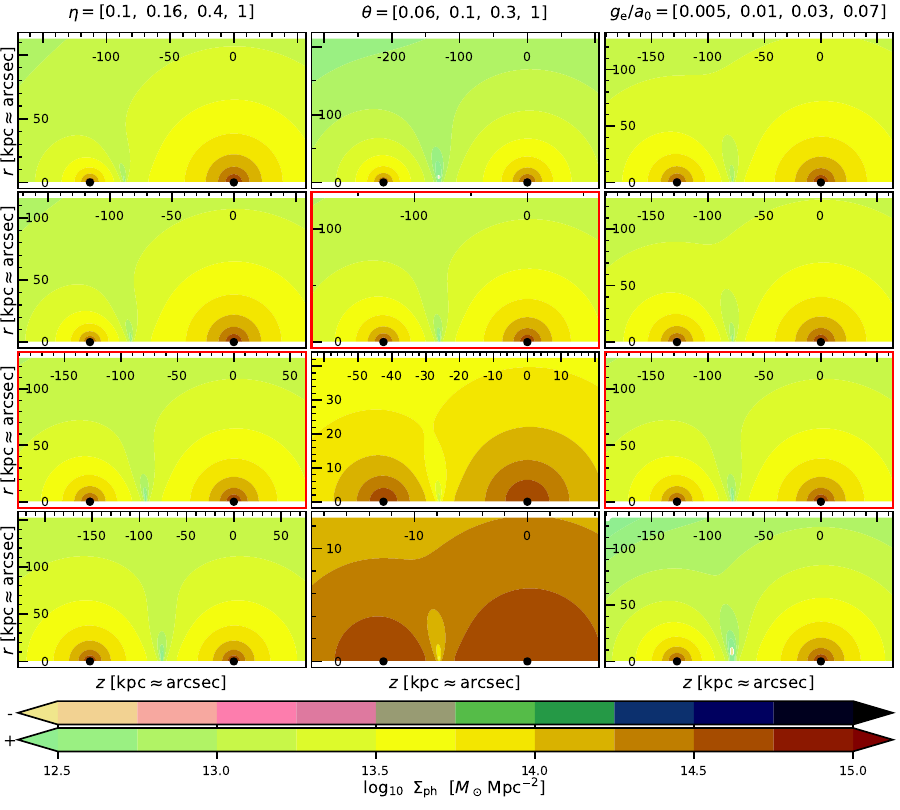}
        \caption{Surface density in the vicinity of the point masses. }
        \label{fig:sigmazoom}
\end{figure*}
\clearpage

\begin{figure*}[b!]
        \centering
        \includegraphics[width=17cm]{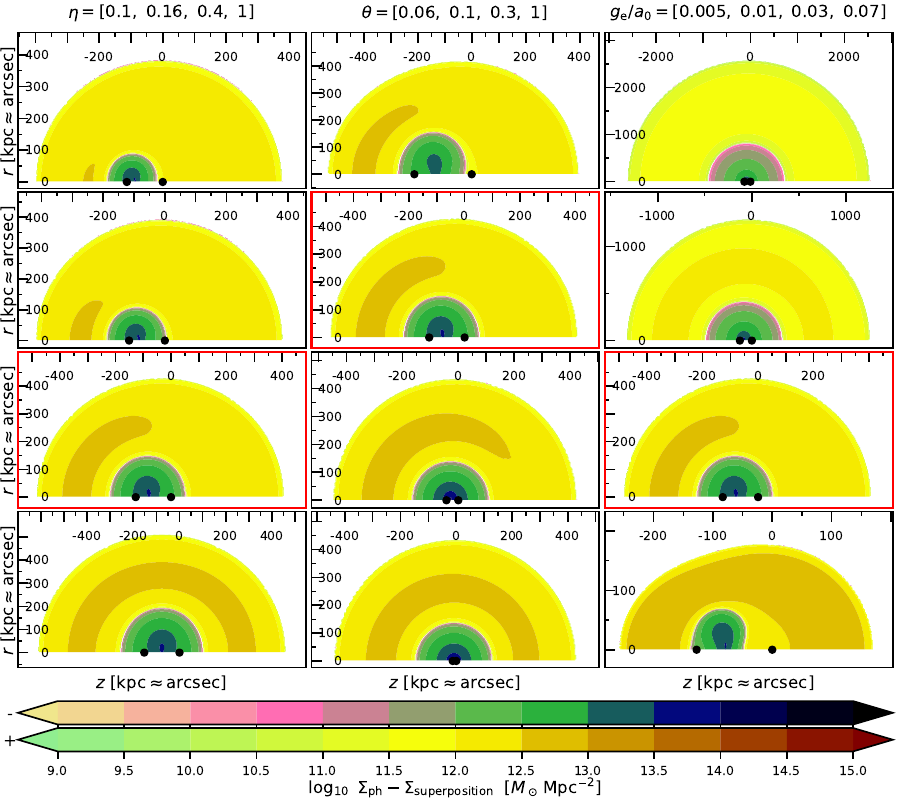}
        \caption{Difference between the true surface density of PDM around two point masses and of the superposition density. The whole halo is depicted.   }
        \label{fig:sigmadiffwide}
\end{figure*}
\clearpage

\begin{figure*}[b!]
        \centering
        \includegraphics[width=17cm]{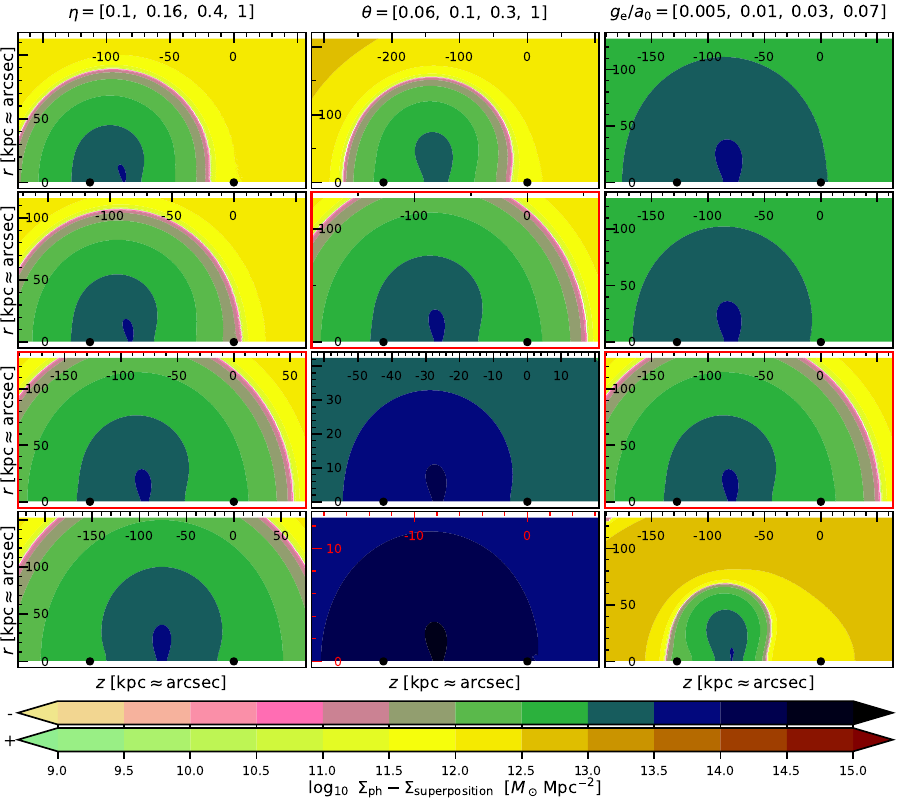}
        \caption{As in \fig{sigmadiffwide}, but only the vicinity of the point masses is shown. }
        \label{fig:sigmadiffzoom}
\end{figure*}
\clearpage

\begin{figure*}[b!]
        \centering
        \includegraphics[width=17cm]{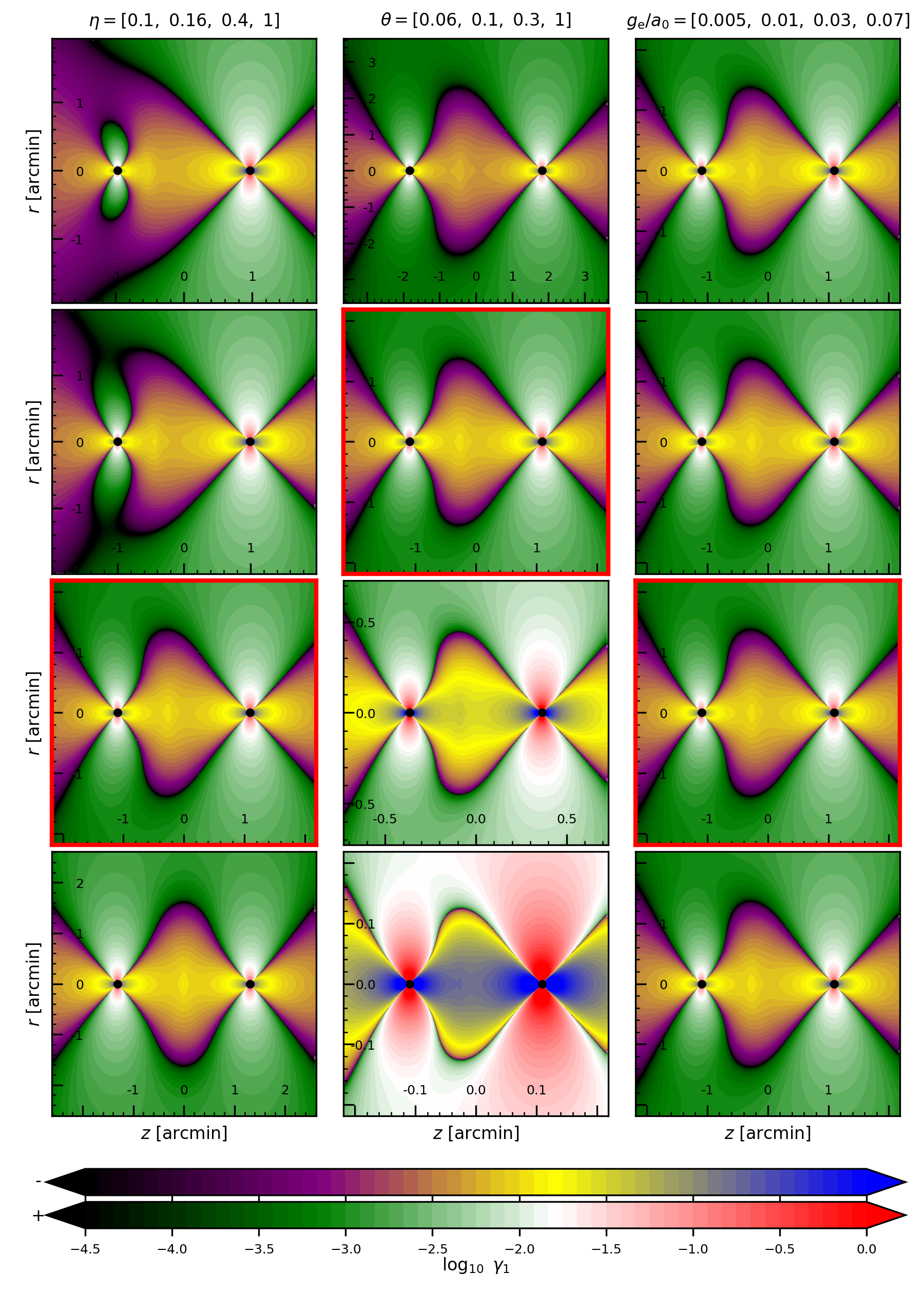}
        \caption{{Real part of the lensing shear. }}
        \label{fig:mapsgam1}
\end{figure*}
\clearpage

\begin{figure*}[b!]
        \centering
        \includegraphics[width=17cm]{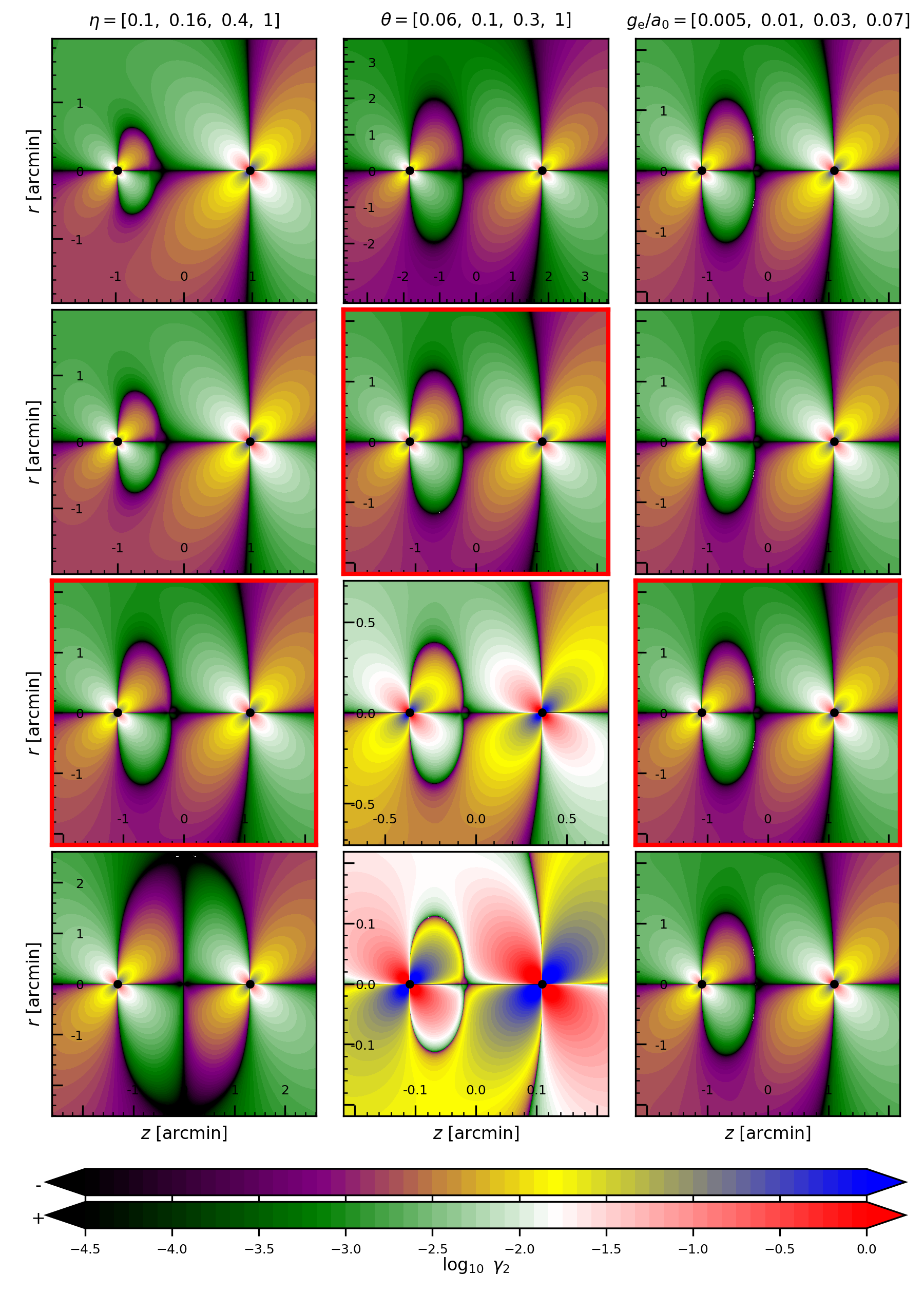}
        \caption{{Imaginary part of the lensing shear.} }
        \label{fig:mapsgam2}
\end{figure*}
\clearpage

\begin{figure*}[b!]
        \centering
        \includegraphics[width=17cm]{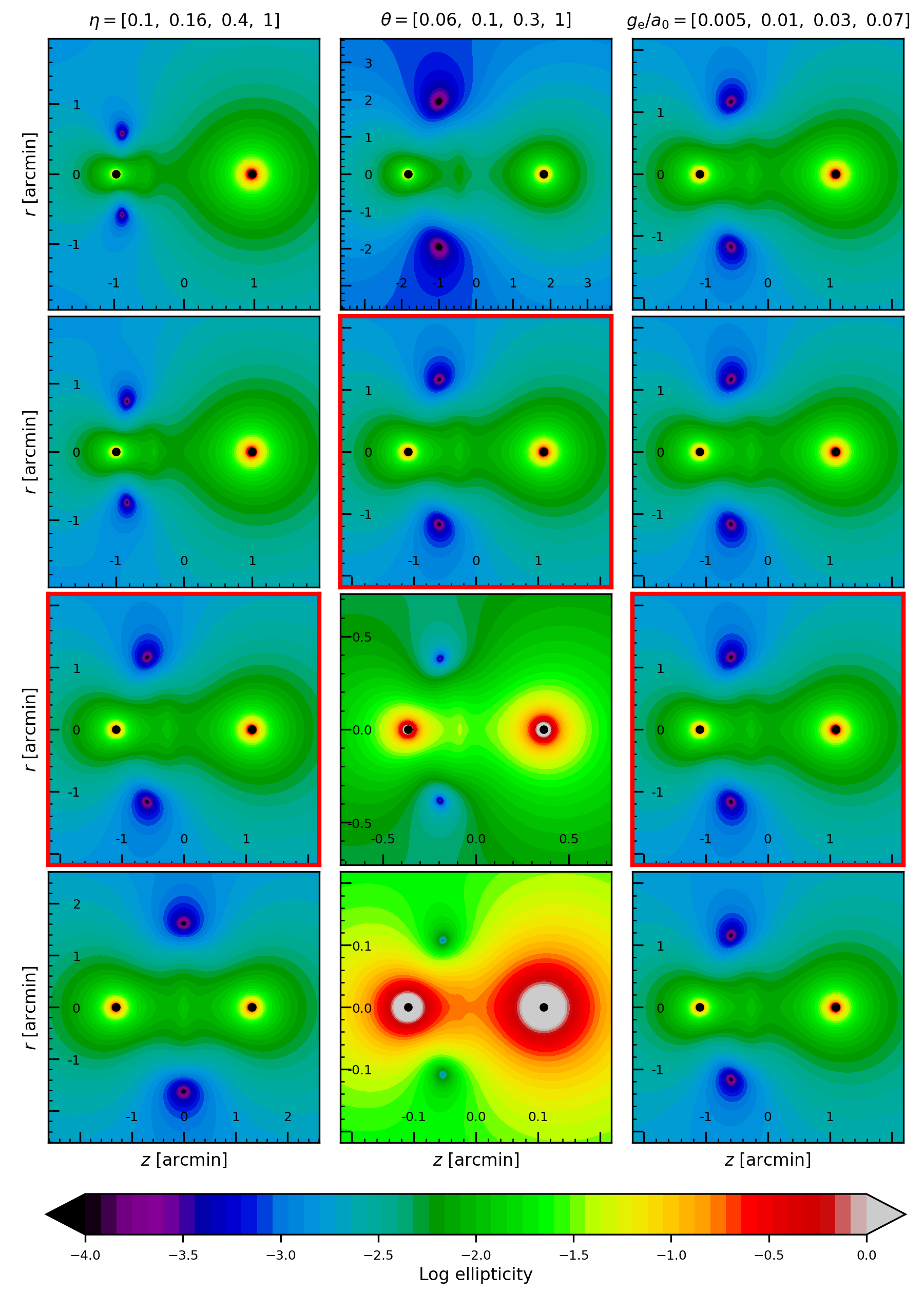}
        \caption{Lensing ellipticity. }
        \label{fig:mapsel}
\end{figure*}
\clearpage

\begin{figure*}[b!]
        \centering
        \includegraphics[width=17cm]{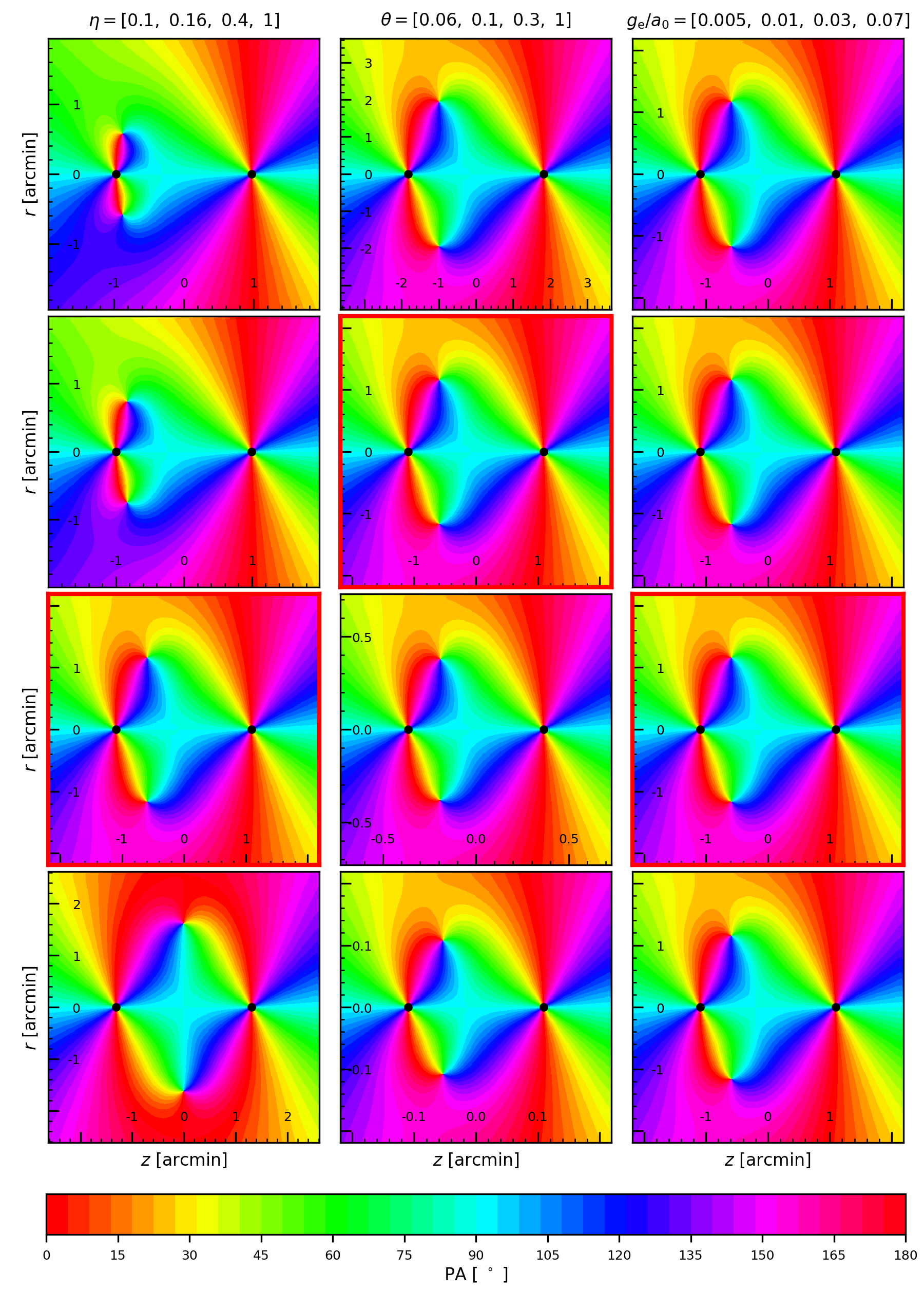}
        \caption{Lensing position angle. }
        \label{fig:mapspa}
\end{figure*}
\clearpage

\begin{figure*}[b!]
        \centering
        \includegraphics[width=17cm]{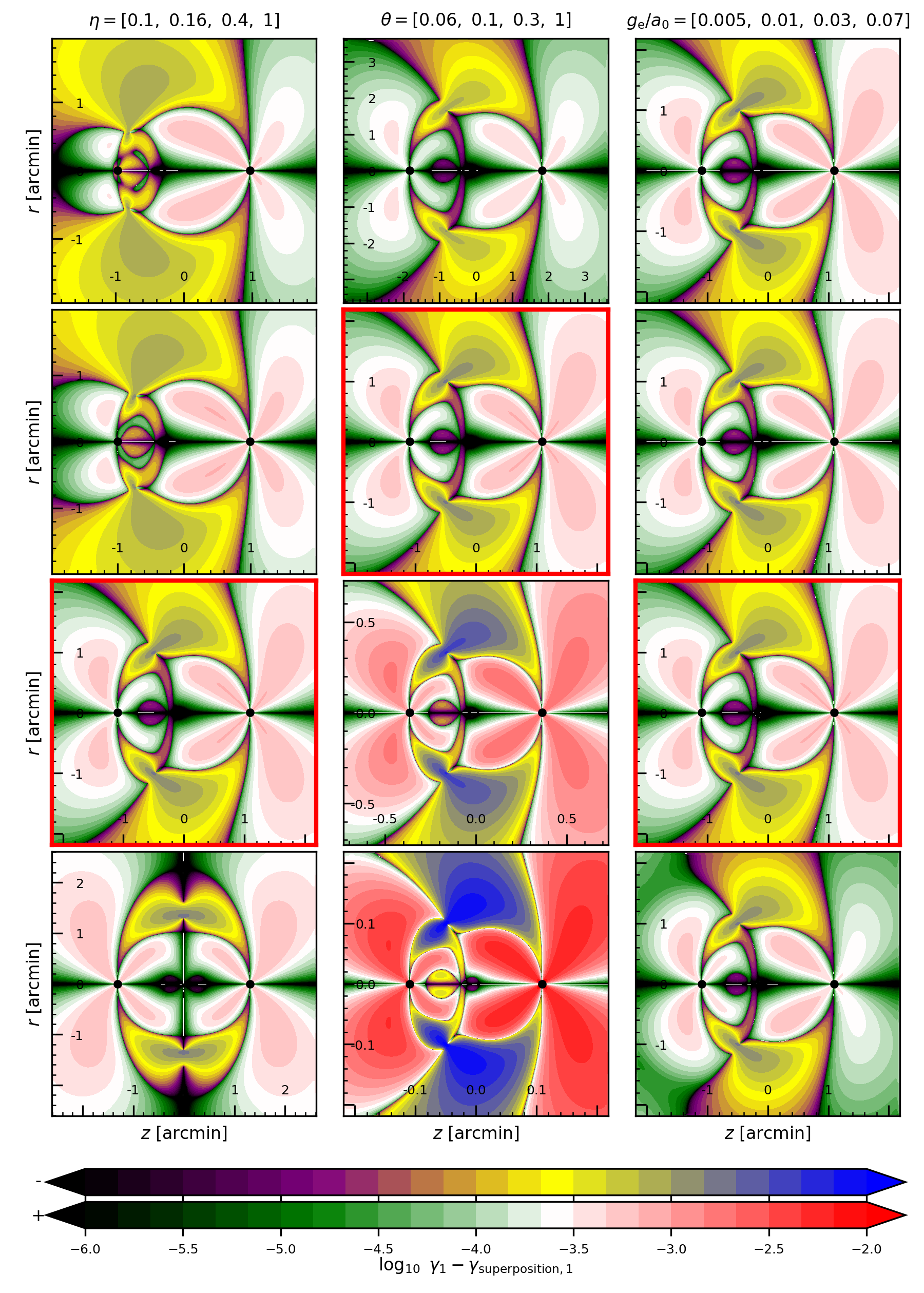}
        \caption{{Difference of the real component of lensing shear produced by the true PDM density and that produced by the superposition density.}}
        \label{fig:mapsgam1diff}
\end{figure*}
\clearpage

\begin{figure*}[b!]
        \centering
        \includegraphics[width=17cm]{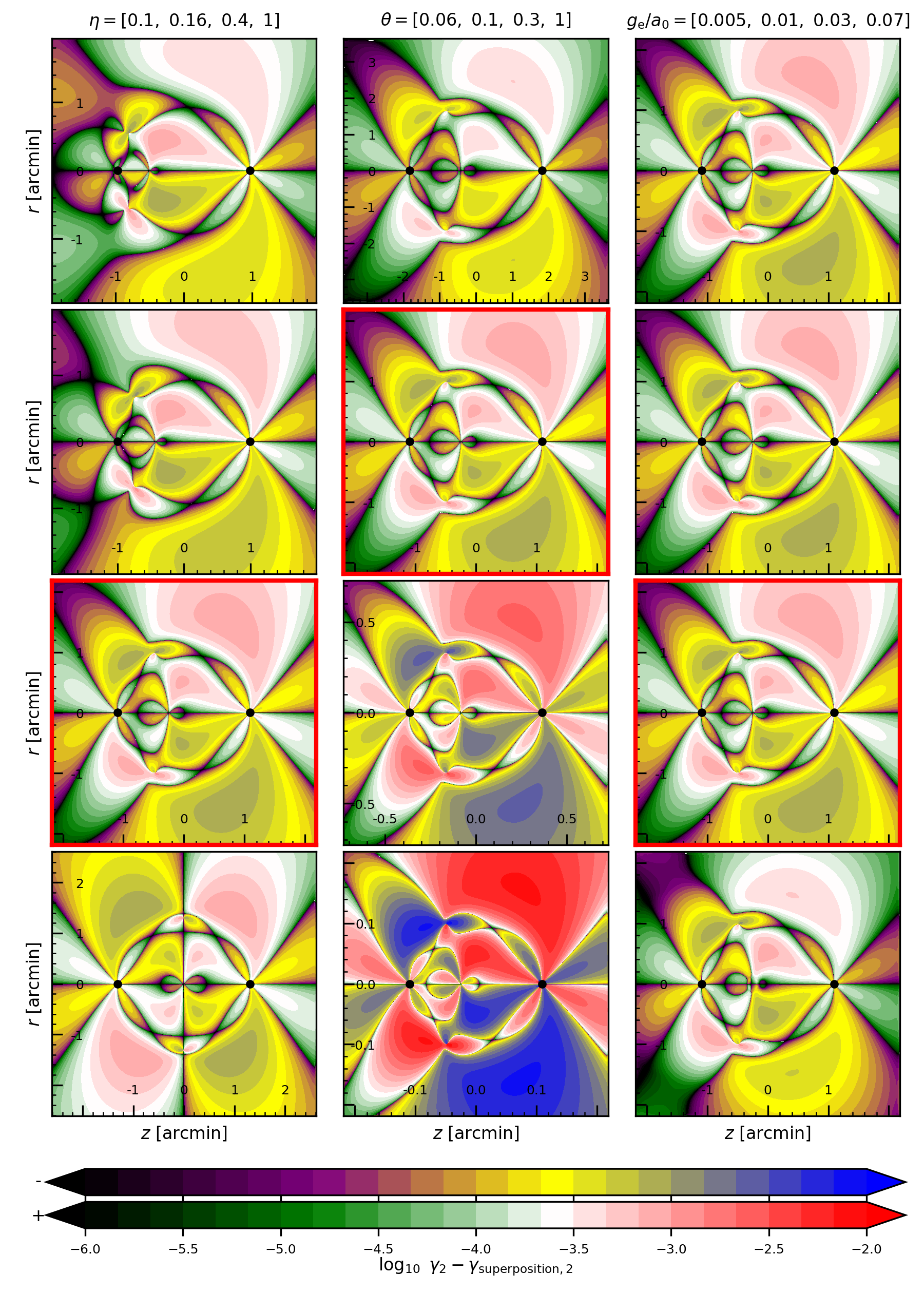}
        \caption{{The same as in \fig{mapsgam2diff} but for the imaginary component of the shear.}}
        \label{fig:mapsgam2diff}
\end{figure*}
\clearpage

\begin{figure*}[b!]
        \centering
        \includegraphics[width=17cm]{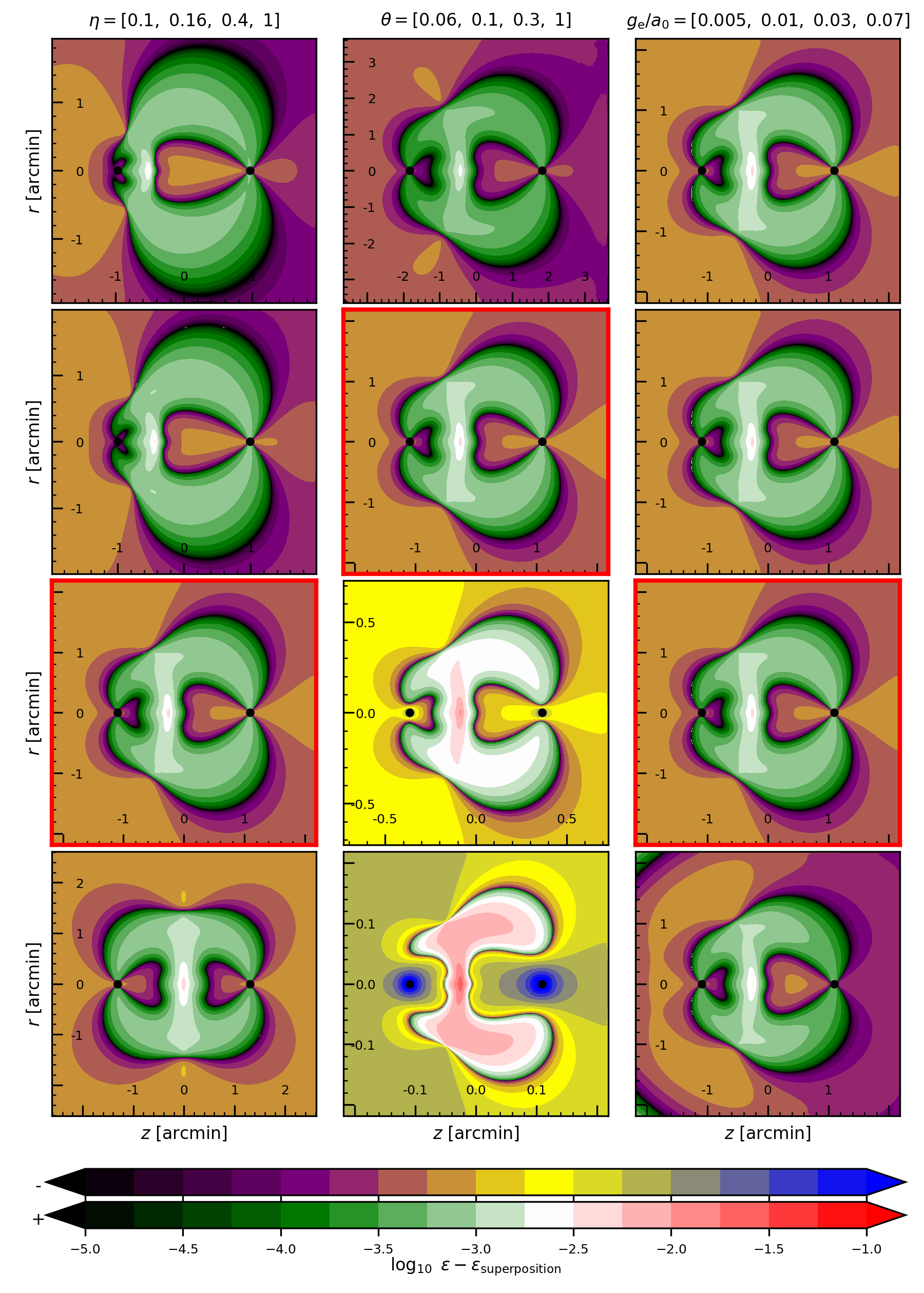}
        \caption{The same as in \fig{mapsgam1diff} but for lensing ellipticity.}
        \label{fig:mapseldiff}
\end{figure*}
\clearpage

\begin{figure*}[b!]
        \centering
        \includegraphics[width=17cm]{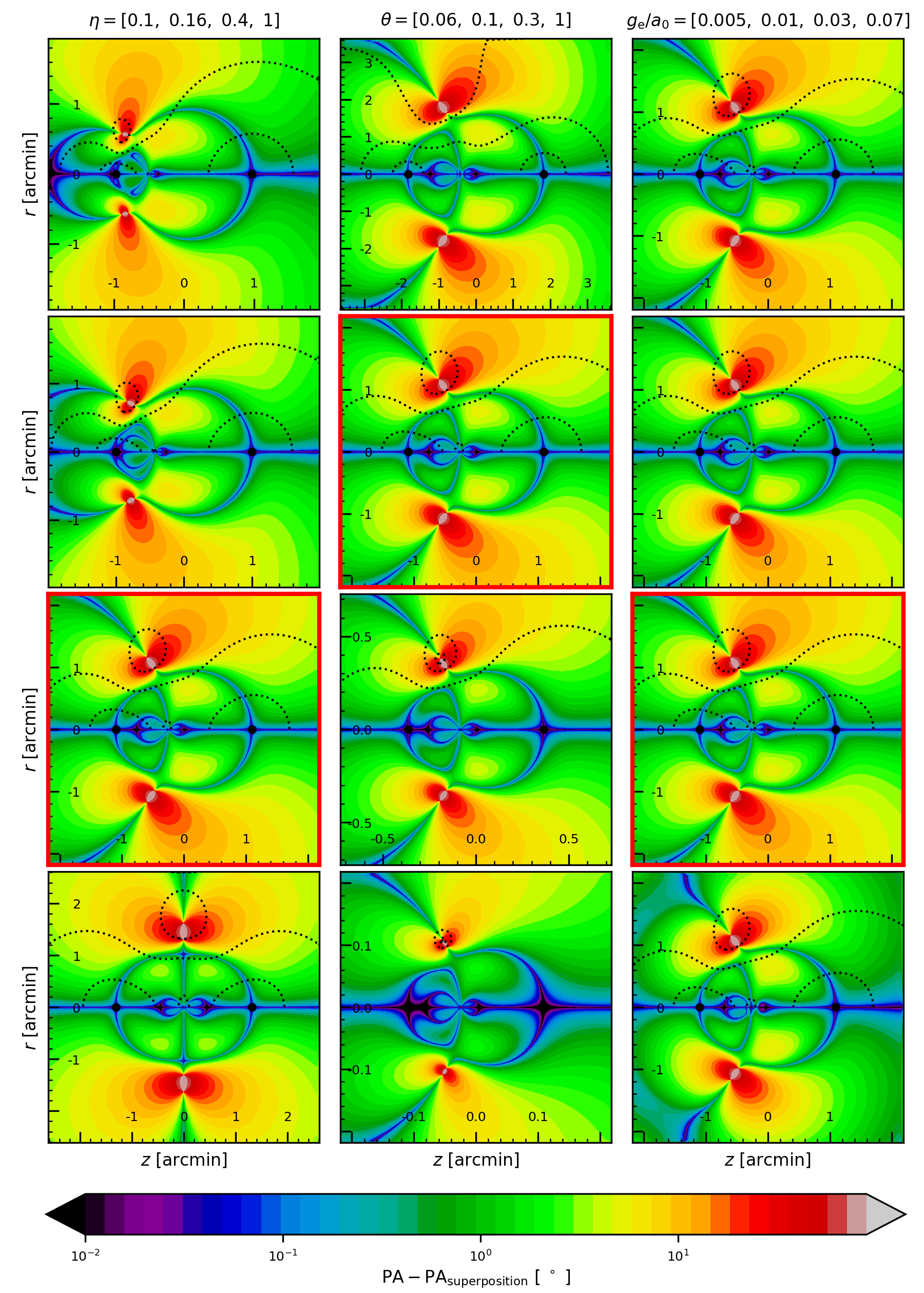}
        \caption{The same as in \fig{mapsgam1diff} but for positional angle. The dotted lines in the upper half of the plots show the contours of the lensing ellipticity of -3,-2.5 and 2, according to \fig{mapsel}.}
        \label{fig:mapspadiff}
\end{figure*}
\clearpage

\begin{figure*}[b!]
        \centering
        \includegraphics[width=17cm]{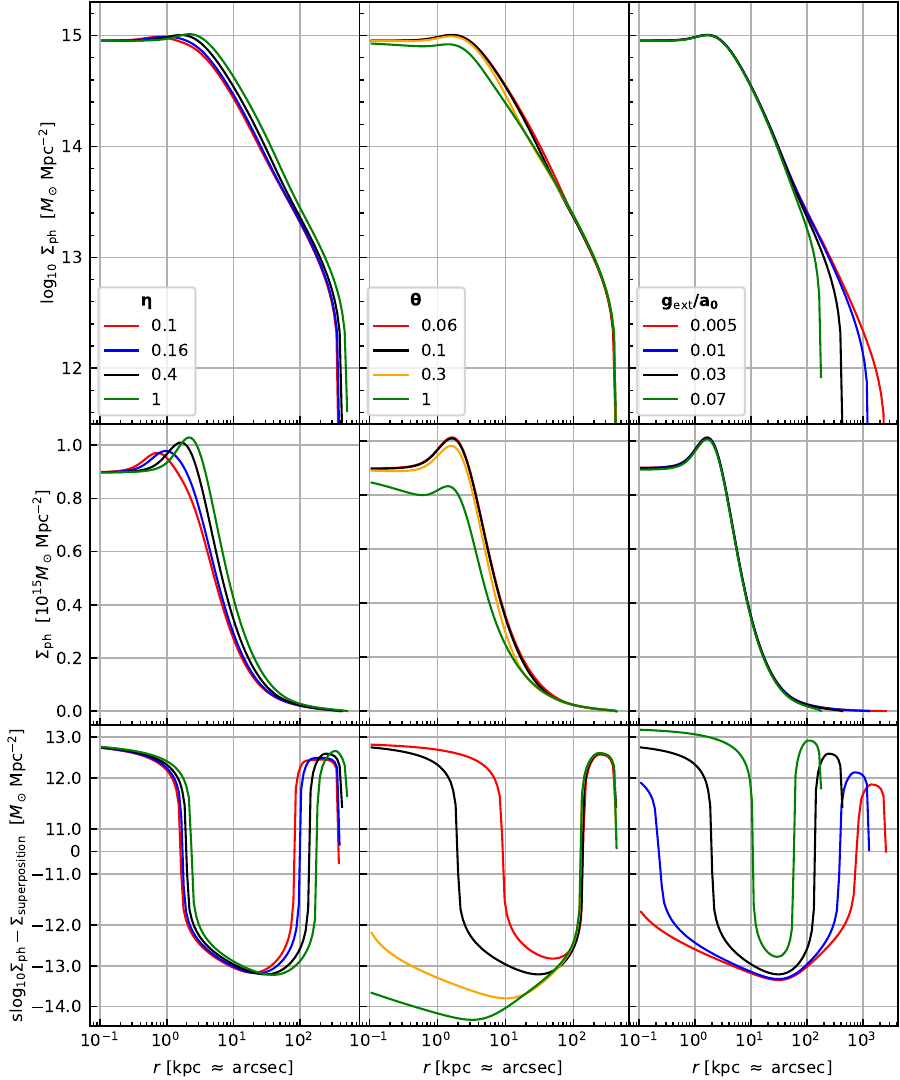}
        \caption{Surface density of the PDM for a pair of point masses observed along their connecting line. In each column, one of the parameters of the fiducial model is varied, as indicated in the legends of the columns.  The black curves always correspond to the fiducial model. Top row: Surface density in the logarithmic scale. Middle row: The same in the linear scale. Bottom row: Difference between the true surface density of the PDM and the superposition surface density. {The symbol ``slog$_{10}$'' stands for the symmetrized decadic logarithm with the linear part between $-10^{11}$ and $10^{11}$.}}
        \label{fig:lossig}
\end{figure*}
\clearpage

\begin{figure*}[b!]
        \centering
        \includegraphics[width=17cm]{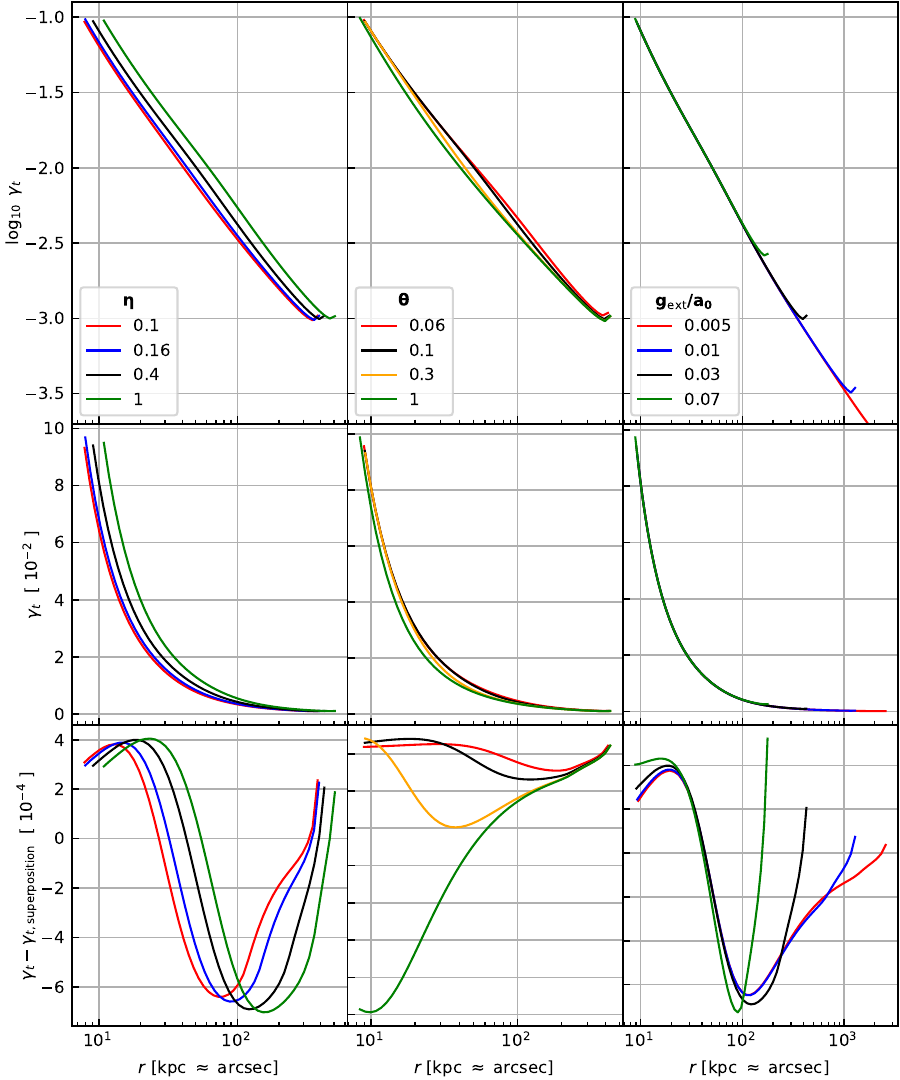}
        \caption{{Tangential shear} for a pair of point masses observed along their connecting line. Top row: {Tangential shear} in the logarithmic scale. Middle row: The same in the linear scale. Bottom row: Difference between the {tangential shear} expected for the true PDM density and the {tangential shear} expected for the superposition density.}
        \label{fig:losell}
\end{figure*}
\clearpage

\onecolumn
\section{Expression for the PDM density around a point mass residing in a homogeneous external field}
\label{app:pm}
\url{(M*exp((gNz^2*(x^2 + y^2 + z^2)^3 + G^2*M^2*x^2 + G^2*M^2*y^2 + G^2*M^2*z^2 - 2*G*M*gNz*z*(x^2 + y^2 + z^2)^(3/2))^(1/4)/(a0^(1/2)*(x^2 + y^2 + z^2)^(3/4)))*(2*gNz^2*z^2*(x^2 + y^2 + z^2)^3 - gNz^2*y^2*(x^2 + y^2 + z^2)^3 - gNz^2*x^2*(x^2 + y^2 + z^2)^3 + 2*G^2*M^2*x^4 + 2*G^2*M^2*y^4 + 2*G^2*M^2*z^4 + 4*G^2*M^2*x^2*y^2 + 4*G^2*M^2*x^2*z^2 + 4*G^2*M^2*y^2*z^2 + G*M*gNz*z*(x^2 + y^2 + z^2)^(5/2) - 5*G*M*gNz*z^3*(x^2 + y^2 + z^2)^(3/2) - 5*G*M*gNz*x^2*z*(x^2 + y^2 + z^2)^(3/2) - 5*G*M*gNz*y^2*z*(x^2 + y^2 + z^2)^(3/2)))/(8*pi*a0^(1/2)*(exp((gNz^2*(x^2 + y^2 + z^2)^3 + G^2*M^2*x^2 + G^2*M^2*y^2 + G^2*M^2*z^2 - 2*G*M*gNz*z*(x^2 + y^2 + z^2)^(3/2))^(1/4)/(a0^(1/2)*(x^2 + y^2 + z^2)^(3/4))) - 1)^2*(x^2 + y^2 + z^2)^(13/4)*(gNz^2*(x^2 + y^2 + z^2)^3 + G^2*M^2*x^2 + G^2*M^2*y^2 + G^2*M^2*z^2 - 2*G*M*gNz*z*(x^2 + y^2 + z^2)^(3/2))^(3/4))}

\section{Expression for the  PDM density around an isolated pair of point masses}
\label{app:tpm}
\url{-(((G*M2)/((d + z)^2 + r^2)^(3/2) + (G*M1)/(r^2 + z^2)^(3/2) - (3*G*M1*z^2)/(r^2 + z^2)^(5/2) - (3*G*M2*(2*d + 2*z)^2)/(4*((d + z)^2 + r^2)^(5/2)))/(exp(-((abs((G*M2*(2*d + 2*z))/(2*((d + z)^2 + r^2)^(3/2)) + (G*M1*z)/(r^2 + z^2)^(3/2))^2 + ((G*M2*r)/((d + z)^2 + r^2)^(3/2) + (G*M1*r)/(r^2 + z^2)^(3/2))^2)^(1/2)/a0)^(1/2)) - 1) + (((G*M2*r)/((d + z)^2 + r^2)^(3/2) + (G*M1*r)/(r^2 + z^2)^(3/2))/(exp(-((abs((G*M2*(2*d + 2*z))/(2*((d + z)^2 + r^2)^(3/2)) + (G*M1*z)/(r^2 + z^2)^(3/2))^2 + ((G*M2*r)/((d + z)^2 + r^2)^(3/2) + (G*M1*r)/(r^2 + z^2)^(3/2))^2)^(1/2)/a0)^(1/2)) - 1) + (r*((G*M2)/((d + z)^2 + r^2)^(3/2) + (G*M1)/(r^2 + z^2)^(3/2) - (3*G*M1*r^2)/(r^2 + z^2)^(5/2) - (3*G*M2*r^2)/((d + z)^2 + r^2)^(5/2)))/(exp(-((abs((G*M2*(2*d + 2*z))/(2*((d + z)^2 + r^2)^(3/2)) + (G*M1*z)/(r^2 + z^2)^(3/2))^2 + ((G*M2*r)/((d + z)^2 + r^2)^(3/2) + (G*M1*r)/(r^2 + z^2)^(3/2))^2)^(1/2)/a0)^(1/2)) - 1) + (r*exp(-((abs((G*M2*(2*d + 2*z))/(2*((d + z)^2 + r^2)^(3/2)) + (G*M1*z)/(r^2 + z^2)^(3/2))^2 + ((G*M2*r)/((d + z)^2 + r^2)^(3/2) + (G*M1*r)/(r^2 + z^2)^(3/2))^2)^(1/2)/a0)^(1/2))*(2*((G*M2*r)/((d + z)^2 + r^2)^(3/2) + (G*M1*r)/(r^2 + z^2)^(3/2))*((G*M2)/((d + z)^2 + r^2)^(3/2) + (G*M1)/(r^2 + z^2)^(3/2) - (3*G*M1*r^2)/(r^2 + z^2)^(5/2) - (3*G*M2*r^2)/((d + z)^2 + r^2)^(5/2)) - 2*abs((G*M2*(2*d + 2*z))/(2*((d + z)^2 + r^2)^(3/2)) + (G*M1*z)/(r^2 + z^2)^(3/2))*sign((G*M2*(2*d + 2*z))/(2*((d + z)^2 + r^2)^(3/2)) + (G*M1*z)/(r^2 + z^2)^(3/2))*((3*G*M1*r*z)/(r^2 + z^2)^(5/2) + (3*G*M2*r*(2*d + 2*z))/(2*((d + z)^2 + r^2)^(5/2))))*((G*M2*r)/((d + z)^2 + r^2)^(3/2) + (G*M1*r)/(r^2 + z^2)^(3/2)))/(4*a0*((abs((G*M2*(2*d + 2*z))/(2*((d + z)^2 + r^2)^(3/2)) + (G*M1*z)/(r^2 + z^2)^(3/2))^2 + ((G*M2*r)/((d + z)^2 + r^2)^(3/2) + (G*M1*r)/(r^2 + z^2)^(3/2))^2)^(1/2)/a0)^(1/2)*(abs((G*M2*(2*d + 2*z))/(2*((d + z)^2 + r^2)^(3/2)) + (G*M1*z)/(r^2 + z^2)^(3/2))^2 + ((G*M2*r)/((d + z)^2 + r^2)^(3/2) + (G*M1*r)/(r^2 + z^2)^(3/2))^2)^(1/2)*(exp(-((abs((G*M2*(2*d + 2*z))/(2*((d + z)^2 + r^2)^(3/2)) + (G*M1*z)/(r^2 + z^2)^(3/2))^2 + ((G*M2*r)/((d + z)^2 + r^2)^(3/2) + (G*M1*r)/(r^2 + z^2)^(3/2))^2)^(1/2)/a0)^(1/2)) - 1)^2))/r - (exp(-((abs((G*M2*(2*d + 2*z))/(2*((d + z)^2 + r^2)^(3/2)) + (G*M1*z)/(r^2 + z^2)^(3/2))^2 + ((G*M2*r)/((d + z)^2 + r^2)^(3/2) + (G*M1*r)/(r^2 + z^2)^(3/2))^2)^(1/2)/a0)^(1/2))*(2*((3*G*M1*r*z)/(r^2 + z^2)^(5/2) + (3*G*M2*r*(2*d + 2*z))/(2*((d + z)^2 + r^2)^(5/2)))*((G*M2*r)/((d + z)^2 + r^2)^(3/2) + (G*M1*r)/(r^2 + z^2)^(3/2)) - 2*abs((G*M2*(2*d + 2*z))/(2*((d + z)^2 + r^2)^(3/2)) + (G*M1*z)/(r^2 + z^2)^(3/2))*sign((G*M2*(2*d + 2*z))/(2*((d + z)^2 + r^2)^(3/2)) + (G*M1*z)/(r^2 + z^2)^(3/2))*((G*M2)/((d + z)^2 + r^2)^(3/2) + (G*M1)/(r^2 + z^2)^(3/2) - (3*G*M1*z^2)/(r^2 + z^2)^(5/2) - (3*G*M2*(2*d + 2*z)^2)/(4*((d + z)^2 + r^2)^(5/2))))*((G*M2*(2*d + 2*z))/(2*((d + z)^2 + r^2)^(3/2)) + (G*M1*z)/(r^2 + z^2)^(3/2)))/(4*a0*((abs((G*M2*(2*d + 2*z))/(2*((d + z)^2 + r^2)^(3/2)) + (G*M1*z)/(r^2 + z^2)^(3/2))^2 + ((G*M2*r)/((d + z)^2 + r^2)^(3/2) + (G*M1*r)/(r^2 + z^2)^(3/2))^2)^(1/2)/a0)^(1/2)*(abs((G*M2*(2*d + 2*z))/(2*((d + z)^2 + r^2)^(3/2)) + (G*M1*z)/(r^2 + z^2)^(3/2))^2 + ((G*M2*r)/((d + z)^2 + r^2)^(3/2) + (G*M1*r)/(r^2 + z^2)^(3/2))^2)^(1/2)*(exp(-((abs((G*M2*(2*d + 2*z))/(2*((d + z)^2 + r^2)^(3/2)) + (G*M1*z)/(r^2 + z^2)^(3/2))^2 + ((G*M2*r)/((d + z)^2 + r^2)^(3/2) + (G*M1*r)/(r^2 + z^2)^(3/2))^2)^(1/2)/a0)^(1/2)) - 1)^2))/(4*G*pi)}

\section{Expression for the  PDM density around two point masses  residing in a homogeneous external field}
\label{app:tpmefe}
\url{-(((G*M1)/(x^2 + y^2 + z^2)^(3/2) + (G*M2)/(2*d*z + d^2 + x^2 + y^2 + z^2)^(3/2) - (3*G*M1*z^2)/(x^2 + y^2 + z^2)^(5/2) - (3*G*M2*(d + z)^2)/(2*d*z + d^2 + x^2 + y^2 + z^2)^(5/2))/(exp(-(((G*M2*(d + z))/(2*d*z + d^2 + x^2 + y^2 + z^2)^(3/2) - gNz + (G*M1*z)/(x^2 + y^2 + z^2)^(3/2))^2 + ((G*M1*x)/(x^2 + y^2 + z^2)^(3/2) - gNx + (G*M2*x)/(2*d*z + d^2 + x^2 + y^2 + z^2)^(3/2))^2 + ((G*M1*y)/(x^2 + y^2 + z^2)^(3/2) - gNy + (G*M2*y)/(2*d*z + d^2 + x^2 + y^2 + z^2)^(3/2))^2)^(1/4)/a0^(1/2)) - 1) + ((G*M1)/(x^2 + y^2 + z^2)^(3/2) + (G*M2)/(2*d*z + d^2 + x^2 + y^2 + z^2)^(3/2) - (3*G*M1*x^2)/(x^2 + y^2 + z^2)^(5/2) - (3*G*M2*x^2)/(2*d*z + d^2 + x^2 + y^2 + z^2)^(5/2))/(exp(-(((G*M2*(d + z))/(2*d*z + d^2 + x^2 + y^2 + z^2)^(3/2) - gNz + (G*M1*z)/(x^2 + y^2 + z^2)^(3/2))^2 + ((G*M1*x)/(x^2 + y^2 + z^2)^(3/2) - gNx + (G*M2*x)/(2*d*z + d^2 + x^2 + y^2 + z^2)^(3/2))^2 + ((G*M1*y)/(x^2 + y^2 + z^2)^(3/2) - gNy + (G*M2*y)/(2*d*z + d^2 + x^2 + y^2 + z^2)^(3/2))^2)^(1/4)/a0^(1/2)) - 1) + ((G*M1)/(x^2 + y^2 + z^2)^(3/2) + (G*M2)/(2*d*z + d^2 + x^2 + y^2 + z^2)^(3/2) - (3*G*M1*y^2)/(x^2 + y^2 + z^2)^(5/2) - (3*G*M2*y^2)/(2*d*z + d^2 + x^2 + y^2 + z^2)^(5/2))/(exp(-(((G*M2*(d + z))/(2*d*z + d^2 + x^2 + y^2 + z^2)^(3/2) - gNz + (G*M1*z)/(x^2 + y^2 + z^2)^(3/2))^2 + ((G*M1*x)/(x^2 + y^2 + z^2)^(3/2) - gNx + (G*M2*x)/(2*d*z + d^2 + x^2 + y^2 + z^2)^(3/2))^2 + ((G*M1*y)/(x^2 + y^2 + z^2)^(3/2) - gNy + (G*M2*y)/(2*d*z + d^2 + x^2 + y^2 + z^2)^(3/2))^2)^(1/4)/a0^(1/2)) - 1) - (exp(-(((G*M2*(d + z))/(2*d*z + d^2 + x^2 + y^2 + z^2)^(3/2) - gNz + (G*M1*z)/(x^2 + y^2 + z^2)^(3/2))^2 + ((G*M1*x)/(x^2 + y^2 + z^2)^(3/2) - gNx + (G*M2*x)/(2*d*z + d^2 + x^2 + y^2 + z^2)^(3/2))^2 + ((G*M1*y)/(x^2 + y^2 + z^2)^(3/2) - gNy + (G*M2*y)/(2*d*z + d^2 + x^2 + y^2 + z^2)^(3/2))^2)^(1/4)/a0^(1/2))*((G*M1*x)/(x^2 + y^2 + z^2)^(3/2) - gNx + (G*M2*x)/(2*d*z + d^2 + x^2 + y^2 + z^2)^(3/2))*(2*((3*G*M1*x*y)/(x^2 + y^2 + z^2)^(5/2) + (3*G*M2*x*y)/(2*d*z + d^2 + x^2 + y^2 + z^2)^(5/2))*((G*M1*y)/(x^2 + y^2 + z^2)^(3/2) - gNy + (G*M2*y)/(2*d*z + d^2 + x^2 + y^2 + z^2)^(3/2)) - 2*((G*M1*x)/(x^2 + y^2 + z^2)^(3/2) - gNx + (G*M2*x)/(2*d*z + d^2 + x^2 + y^2 + z^2)^(3/2))*((G*M1)/(x^2 + y^2 + z^2)^(3/2) + (G*M2)/(2*d*z + d^2 + x^2 + y^2 + z^2)^(3/2) - (3*G*M1*x^2)/(x^2 + y^2 + z^2)^(5/2) - (3*G*M2*x^2)/(2*d*z + d^2 + x^2 + y^2 + z^2)^(5/2)) + 2*((3*G*M1*x*z)/(x^2 + y^2 + z^2)^(5/2) + (3*G*M2*x*(2*d + 2*z))/(2*(2*d*z + d^2 + x^2 + y^2 + z^2)^(5/2)))*((G*M2*(d + z))/(2*d*z + d^2 + x^2 + y^2 + z^2)^(3/2) - gNz + (G*M1*z)/(x^2 + y^2 + z^2)^(3/2))))/(4*a0^(1/2)*(exp(-(((G*M2*(d + z))/(2*d*z + d^2 + x^2 + y^2 + z^2)^(3/2) - gNz + (G*M1*z)/(x^2 + y^2 + z^2)^(3/2))^2 + ((G*M1*x)/(x^2 + y^2 + z^2)^(3/2) - gNx + (G*M2*x)/(2*d*z + d^2 + x^2 + y^2 + z^2)^(3/2))^2 + ((G*M1*y)/(x^2 + y^2 + z^2)^(3/2) - gNy + (G*M2*y)/(2*d*z + d^2 + x^2 + y^2 + z^2)^(3/2))^2)^(1/4)/a0^(1/2)) - 1)^2*(((G*M2*(d + z))/(2*d*z + d^2 + x^2 + y^2 + z^2)^(3/2) - gNz + (G*M1*z)/(x^2 + y^2 + z^2)^(3/2))^2 + ((G*M1*x)/(x^2 + y^2 + z^2)^(3/2) - gNx + (G*M2*x)/(2*d*z + d^2 + x^2 + y^2 + z^2)^(3/2))^2 + ((G*M1*y)/(x^2 + y^2 + z^2)^(3/2) - gNy + (G*M2*y)/(2*d*z + d^2 + x^2 + y^2 + z^2)^(3/2))^2)^(3/4)) - (exp(-(((G*M2*(d + z))/(2*d*z + d^2 + x^2 + y^2 + z^2)^(3/2) - gNz + (G*M1*z)/(x^2 + y^2 + z^2)^(3/2))^2 + ((G*M1*x)/(x^2 + y^2 + z^2)^(3/2) - gNx + (G*M2*x)/(2*d*z + d^2 + x^2 + y^2 + z^2)^(3/2))^2 + ((G*M1*y)/(x^2 + y^2 + z^2)^(3/2) - gNy + (G*M2*y)/(2*d*z + d^2 + x^2 + y^2 + z^2)^(3/2))^2)^(1/4)/a0^(1/2))*((G*M1*y)/(x^2 + y^2 + z^2)^(3/2) - gNy + (G*M2*y)/(2*d*z + d^2 + x^2 + y^2 + z^2)^(3/2))*(2*((3*G*M1*x*y)/(x^2 + y^2 + z^2)^(5/2) + (3*G*M2*x*y)/(2*d*z + d^2 + x^2 + y^2 + z^2)^(5/2))*((G*M1*x)/(x^2 + y^2 + z^2)^(3/2) - gNx + (G*M2*x)/(2*d*z + d^2 + x^2 + y^2 + z^2)^(3/2)) - 2*((G*M1*y)/(x^2 + y^2 + z^2)^(3/2) - gNy + (G*M2*y)/(2*d*z + d^2 + x^2 + y^2 + z^2)^(3/2))*((G*M1)/(x^2 + y^2 + z^2)^(3/2) + (G*M2)/(2*d*z + d^2 + x^2 + y^2 + z^2)^(3/2) - (3*G*M1*y^2)/(x^2 + y^2 + z^2)^(5/2) - (3*G*M2*y^2)/(2*d*z + d^2 + x^2 + y^2 + z^2)^(5/2)) + 2*((3*G*M1*y*z)/(x^2 + y^2 + z^2)^(5/2) + (3*G*M2*y*(2*d + 2*z))/(2*(2*d*z + d^2 + x^2 + y^2 + z^2)^(5/2)))*((G*M2*(d + z))/(2*d*z + d^2 + x^2 + y^2 + z^2)^(3/2) - gNz + (G*M1*z)/(x^2 + y^2 + z^2)^(3/2))))/(4*a0^(1/2)*(exp(-(((G*M2*(d + z))/(2*d*z + d^2 + x^2 + y^2 + z^2)^(3/2) - gNz + (G*M1*z)/(x^2 + y^2 + z^2)^(3/2))^2 + ((G*M1*x)/(x^2 + y^2 + z^2)^(3/2) - gNx + (G*M2*x)/(2*d*z + d^2 + x^2 + y^2 + z^2)^(3/2))^2 + ((G*M1*y)/(x^2 + y^2 + z^2)^(3/2) - gNy + (G*M2*y)/(2*d*z + d^2 + x^2 + y^2 + z^2)^(3/2))^2)^(1/4)/a0^(1/2)) - 1)^2*(((G*M2*(d + z))/(2*d*z + d^2 + x^2 + y^2 + z^2)^(3/2) - gNz + (G*M1*z)/(x^2 + y^2 + z^2)^(3/2))^2 + ((G*M1*x)/(x^2 + y^2 + z^2)^(3/2) - gNx + (G*M2*x)/(2*d*z + d^2 + x^2 + y^2 + z^2)^(3/2))^2 + ((G*M1*y)/(x^2 + y^2 + z^2)^(3/2) - gNy + (G*M2*y)/(2*d*z + d^2 + x^2 + y^2 + z^2)^(3/2))^2)^(3/4)) - (exp(-(((G*M2*(d + z))/(2*d*z + d^2 + x^2 + y^2 + z^2)^(3/2) - gNz + (G*M1*z)/(x^2 + y^2 + z^2)^(3/2))^2 + ((G*M1*x)/(x^2 + y^2 + z^2)^(3/2) - gNx + (G*M2*x)/(2*d*z + d^2 + x^2 + y^2 + z^2)^(3/2))^2 + ((G*M1*y)/(x^2 + y^2 + z^2)^(3/2) - gNy + (G*M2*y)/(2*d*z + d^2 + x^2 + y^2 + z^2)^(3/2))^2)^(1/4)/a0^(1/2))*(2*((3*G*M1*x*z)/(x^2 + y^2 + z^2)^(5/2) + (3*G*M2*x*(2*d + 2*z))/(2*(2*d*z + d^2 + x^2 + y^2 + z^2)^(5/2)))*((G*M1*x)/(x^2 + y^2 + z^2)^(3/2) - gNx + (G*M2*x)/(2*d*z + d^2 + x^2 + y^2 + z^2)^(3/2)) + 2*((3*G*M1*y*z)/(x^2 + y^2 + z^2)^(5/2) + (3*G*M2*y*(2*d + 2*z))/(2*(2*d*z + d^2 + x^2 + y^2 + z^2)^(5/2)))*((G*M1*y)/(x^2 + y^2 + z^2)^(3/2) - gNy + (G*M2*y)/(2*d*z + d^2 + x^2 + y^2 + z^2)^(3/2)) - 2*((G*M2*(d + z))/(2*d*z + d^2 + x^2 + y^2 + z^2)^(3/2) - gNz + (G*M1*z)/(x^2 + y^2 + z^2)^(3/2))*((G*M1)/(x^2 + y^2 + z^2)^(3/2) + (G*M2)/(2*d*z + d^2 + x^2 + y^2 + z^2)^(3/2) - (3*G*M1*z^2)/(x^2 + y^2 + z^2)^(5/2) - (3*G*M2*(d + z)^2)/(2*d*z + d^2 + x^2 + y^2 + z^2)^(5/2)))*((G*M2*(d + z))/(2*d*z + d^2 + x^2 + y^2 + z^2)^(3/2) - gNz + (G*M1*z)/(x^2 + y^2 + z^2)^(3/2)))/(4*a0^(1/2)*(exp(-(((G*M2*(d + z))/(2*d*z + d^2 + x^2 + y^2 + z^2)^(3/2) - gNz + (G*M1*z)/(x^2 + y^2 + z^2)^(3/2))^2 + ((G*M1*x)/(x^2 + y^2 + z^2)^(3/2) - gNx + (G*M2*x)/(2*d*z + d^2 + x^2 + y^2 + z^2)^(3/2))^2 + ((G*M1*y)/(x^2 + y^2 + z^2)^(3/2) - gNy + (G*M2*y)/(2*d*z + d^2 + x^2 + y^2 + z^2)^(3/2))^2)^(1/4)/a0^(1/2)) - 1)^2*(((G*M2*(d + z))/(2*d*z + d^2 + x^2 + y^2 + z^2)^(3/2) - gNz + (G*M1*z)/(x^2 + y^2 + z^2)^(3/2))^2 + ((G*M1*x)/(x^2 + y^2 + z^2)^(3/2) - gNx + (G*M2*x)/(2*d*z + d^2 + x^2 + y^2 + z^2)^(3/2))^2 + ((G*M1*y)/(x^2 + y^2 + z^2)^(3/2) - gNy + (G*M2*y)/(2*d*z + d^2 + x^2 + y^2 + z^2)^(3/2))^2)^(3/4)))/(4*G*pi)}

\end{appendix}

\end{document}